\documentclass[revtex4,iop,numberedappendix]{emulateapj}
\pdfoutput=1

\usepackage{amsmath}
\usepackage{color}
\usepackage{xspace}
\usepackage{hyperref}
\usepackage{textcomp} % use for the \textminus symbol
\usepackage{acronym}

\newcommand{\ee}[1]{\!\times\!10^{#1}}
\newcommand{\gws}{gravitational waves\xspace}
\newcommand{\gw}{gravitational wave\xspace}
\newcommand{\ghw}{gravitational-wave\xspace}
\newcommand{\geo}{GEO\,600~}

\newcommand{\F}{\ensuremath{\mathcal{F}}}
\newcommand{\G}{\ensuremath{\mathcal{G}}}
\newcommand{\FG}{\F/\G}

\newcommand{\frot}{\ensuremath{f_{\rm rot}}\xspace}
\newcommand{\hl}[1]{{\color{black}#1}}
\newcommand{\tmin}{\textminus}

\newcommand{\NPULSARS}{\hl{200}\xspace}       % number of pulsars in the search
\newcommand{\NHIGHVALUE}{\hl{11}\xspace}      % number of high value targets using multiple pipelines
\newcommand{\NLOWVALUE}{\hl{189}\xspace}      % number of targets just using the Bayesian pipeline
\newcommand{\NBELOWSDWORD}{\hl{eight}\xspace}
\newcommand{\NPREVIOUS}{\hl{122}\xspace}      % number over pulsars overlapping with previous analyses
\newcommand{\NNEW}{\hl{78}\xspace}            % number of pulsars new to this analysis
\newcommand{\LHODUTYFACTOR}{\hl{60\%}\xspace} % LHO duty factor
\newcommand{\LLODUTYFACTOR}{\hl{51\%}\xspace} % LLO duty factor
\newcommand{\LHOOBSTIME}{\hl{78}\xspace}      % LHO science mode observing time (days)
\newcommand{\LLOOBSTIME}{\hl{66}\xspace}      % LLO science mode observing time (days)
\newcommand{\WITHINFACTORTEN}{\hl{32}\xspace} % pulsars above, but within x10 of spin-down limit
\newcommand{\MINHZEROPULSAR}{\hl{PSR\,J1918\tmin0642}\xspace}   % pulsar with minimum h0 upper limit
\newcommand{\MINHZERO}{\hl{$1.6\ee{-26}$}\xspace}          % minimum h0 upper limit
\newcommand{\MINEQPULSAR}{\hl{J0636+5129}\xspace}   % pulsar with minimum ellipticity/Q22 value
\newcommand{\MINELL}{\hl{$1.3\ee{-8}$}\xspace}          % minimum ellipticity
\newcommand{\MINQ}{\hl{$9.7\ee{29}$}\,kg\,m$^2$\xspace} % minimum Q22
\newcommand{\MINEQDIST}{\hl{$\sim 0.20$\,kpc}\xspace}   % distance to minimum ellipticity pulsar

\begin{document}


\title{First search for gravitational waves from known pulsars with Advanced LIGO}
\shorttitle{Gravitational waves from known pulsars}

% includes LVC August 2016 author list https://dcc.ligo.org/LIGO-M1600213

\ifdefined\shortauthor
 
% version to compile if not wanting to include the full author list 

\author{The LIGO Scientific Collaboration}
\author{The Virgo Collaboration}

% pulsar astronomer authors
\author{S.~Buchner\altaffilmark{1,2},  % South Africa
I.~Cognard\altaffilmark{3,4},          % Nancay
A.~Corongiu\altaffilmark{5},           % Cagliari group
P.~C.~C.~Freire\altaffilmark{6},       % Bonn group
L.~Guillemot\altaffilmark{3,4},        % Nancay group
G.~B.~Hobbs\altaffilmark{7},           % ATNF group
M.~Kerr\altaffilmark{7},               % ATNF group
A.~G.~Lyne\altaffilmark{8},            % Jodrell Bank
A.~Possenti\altaffilmark{5},           % Cagliari group
A.~Ridolfi\altaffilmark{6},            % Bonn group
R.~M.~Shannon\altaffilmark{9,10},      % ATNF group
B.~W.~Stappers\altaffilmark{8},        % Jodrell Bank
and P.~Weltevrede\altaffilmark{8}      % Jodrell Bank
} 

\altaffiltext{1}{Square Kilometer Array South Africa, The Park, Park Road, Pinelands, Cape Town 7405, South 
Africa}
\altaffiltext{2}{Hartebeesthoek Radio Astronomy Observatory, PO Box 443, Krugersdorp, 1740, South Africa}
\altaffiltext{3}{Laboratoire de Physique et Chimie de l'Environnement et de l'Espace, LPC2E, 
CNRS-Universit\'{e} d'Orl\'{e}ans, F-45071 Orl\'{e}ans, France}
\altaffiltext{4}{Station de Radioastronomie de Nan\c{c}ay, Observatoire de Paris, CNRS/INSU, F-18330 
Nan\c{c}ay, France}
\altaffiltext{5}{INAF - Osservatorio Astronomico di Cagliari, via della Scienza 5, 09047 Selargius, Italy}
\altaffiltext{6}{Max-Planck-Institut f\"{u}r Radioastronomie MPIfR, Auf dem H\"{u}gel 
69, D-53121 Bonn, Germany}
\altaffiltext{7}{CSIRO Astronomy and Space Science, Australia Telescope National Facility, Box 76 
Epping, NSW, 1710, Australia}
\altaffiltext{8}{Jodrell Bank Centre for Astrophysics, School of Physics and Astronomy, University 
of Manchester, Manchester M13 9PL, UK}
\altaffiltext{9}{CSIRO Astronomy and Space Science, Australia Telescope National Facility, Box 76 
Epping, NSW, 1710, Australia}
\altaffiltext{10}{International Centre for Radio Astronomy Research, Curtin University, Bentley, WA 
6102, Australia}

\else

\author{%
B.~P.~Abbott,\altaffilmark{1}  %benjamin.abbott
R.~Abbott,\altaffilmark{1}  %rich.abbott
T.~D.~Abbott,\altaffilmark{2}  %thomas.abbott
M.~R.~Abernathy,\altaffilmark{3}  %matthew.abernathy
F.~Acernese,\altaffilmark{4,5} %fausto.acernese
K.~Ackley,\altaffilmark{6}  %kendall.ackley
C.~Adams,\altaffilmark{7}  %carl.adams
T.~Adams,\altaffilmark{8} %thomas.adams
P.~Addesso,\altaffilmark{9}  %paolo.addesso
R.~X.~Adhikari,\altaffilmark{1}
% ...
}  %john.zweizig

\medskip
\affiliation {${}^{*}$Deceased, March 2016. 
\\
{(LIGO Scientific Collaboration and Virgo Collaboration)}%
}% 
\medskip

\altaffiltext {1}{LIGO, California Institute of Technology, Pasadena, CA 91125, USA }
\altaffiltext {2}{Louisiana State University, Baton Rouge, LA 70803, USA }
\altaffiltext {3}{American University, Washington, D.C. 20016, USA }
\altaffiltext {4}{Universit\`a di Salerno, Fisciano, I-84084 Salerno, Italy }
\altaffiltext {5}{INFN, Sezione di Napoli, Complesso Universitario di Monte S.Angelo, I-80126 Napoli, Italy }
\altaffiltext {6}{University of Florida, Gainesville, FL 32611, USA }
\altaffiltext {7}{LIGO Livingston Observatory, Livingston, LA 70754, USA }
\altaffiltext {8}{Laboratoire d'Annecy-le-Vieux de Physique des Particules (LAPP), Universit\'e Savoie Mont 
Blanc, CNRS/IN2P3, F-74941 Annecy-le-Vieux, France }
\altaffiltext {9}{University of Sannio at Benevento, I-82100 Benevento, Italy and INFN, Sezione di Napoli, 
I-80100 Napoli, Italy }
\altaffiltext {10}{Albert-Einstein-Institut, Max-Planck-Institut f\"ur Gravi\-ta\-tions\-physik, D-30167 
Hannover, Germany }
\altaffiltext {11}{Nikhef, Science Park, 1098 XG Amsterdam, The Netherlands }
\altaffiltext {12}{LIGO, Massachusetts Institute of Technology, Cambridge, MA 02139, USA }
\altaffiltext {13}{Instituto Nacional de Pesquisas Espaciais, 12227-010 S\~{a}o Jos\'{e} dos Campos, S\~{a}o 
Paulo, Brazil }
\altaffiltext {14}{INFN, Gran Sasso Science Institute, I-67100 L'Aquila, Italy }
\altaffiltext {15}{INFN, Sezione di Roma Tor Vergata, I-00133 Roma, Italy }
\altaffiltext {16}{Inter-University Centre for Astronomy and Astrophysics, Pune 411007, India }
\altaffiltext {17}{International Centre for Theoretical Sciences, Tata Institute of Fundamental Research, 
Bengaluru 560089, India }
\altaffiltext {18}{University of Wisconsin-Milwaukee, Milwaukee, WI 53201, USA }
\altaffiltext {19}{Leibniz Universit\"at Hannover, D-30167 Hannover, Germany }
\altaffiltext {20}{Universit\`a di Pisa, I-56127 Pisa, Italy }
\altaffiltext {21}{INFN, Sezione di Pisa, I-56127 Pisa, Italy }
\altaffiltext {22}{Australian National University, Canberra, Australian Capital Territory 0200, Australia }
\altaffiltext {23}{California State University Fullerton, Fullerton, CA 92831, USA }
\altaffiltext {24}{LAL, Univ. Paris-Sud, CNRS/IN2P3, Universit\'e Paris-Saclay, F-91898 Orsay, France }
\altaffiltext {25}{Chennai Mathematical Institute, Chennai 603103, India }
\altaffiltext {26}{Universit\`a di Roma Tor Vergata, I-00133 Roma, Italy }
\altaffiltext {27}{Universit\"at Hamburg, D-22761 Hamburg, Germany }
\altaffiltext {28}{INFN, Sezione di Roma, I-00185 Roma, Italy }
\altaffiltext {29}{Albert-Einstein-Institut, Max-Planck-Institut f\"ur Gravitations\-physik, D-14476 
Potsdam-Golm, Germany }
\altaffiltext {30}{APC, AstroParticule et Cosmologie, Universit\'e Paris Diderot, CNRS/IN2P3, CEA/Irfu, 
Observatoire de Paris, Sorbonne Paris Cit\'e, F-75205 Paris Cedex 13, France }
\altaffiltext {31}{West Virginia University, Morgantown, WV 26506, USA }
\altaffiltext {32}{Universit\`a di Perugia, I-06123 Perugia, Italy }
\altaffiltext {33}{INFN, Sezione di Perugia, I-06123 Perugia, Italy }
\altaffiltext {34}{European Gravitational Observatory (EGO), I-56021 Cascina, Pisa, Italy }
\altaffiltext {35}{Syracuse University, Syracuse, NY 13244, USA }
\altaffiltext {36}{SUPA, University of Glasgow, Glasgow G12 8QQ, UK }
\altaffiltext {37}{LIGO Hanford Observatory, Richland, WA 99352, USA }
\altaffiltext {38}{Wigner RCP, RMKI, H-1121 Budapest, Konkoly Thege Mikl\'os \'ut 29-33, Hungary }
\altaffiltext {39}{Columbia University, New York, NY 10027, USA }
\altaffiltext {40}{Stanford University, Stanford, CA 94305, USA }
\altaffiltext {41}{Universit\`a di Padova, Dipartimento di Fisica e Astronomia, I-35131 Padova, Italy }
\altaffiltext {42}{INFN, Sezione di Padova, I-35131 Padova, Italy }
\altaffiltext {43}{Nicolaus Copernicus Astronomical Center, Polish Academy of Sciences, 00-716, Warsaw, 
Poland }
\altaffiltext {44}{Center for Relativistic Astrophysics and School of Physics, Georgia Institute of 
Technology, Atlanta, GA 30332, USA }
\altaffiltext {45}{University of Birmingham, Birmingham B15 2TT, UK }
\altaffiltext {46}{Universit\`a degli Studi di Genova, I-16146 Genova, Italy }
\altaffiltext {47}{INFN, Sezione di Genova, I-16146 Genova, Italy }
\altaffiltext {48}{RRCAT, Indore MP 452013, India }
\altaffiltext {49}{Faculty of Physics, Lomonosov Moscow State University, Moscow 119991, Russia }
\altaffiltext {50}{SUPA, University of the West of Scotland, Paisley PA1 2BE, UK }
\altaffiltext {51}{Caltech CaRT, Pasadena, CA 91125, USA }
\altaffiltext {52}{University of Western Australia, Crawley, Western Australia 6009, Australia }
\altaffiltext {53}{Department of Astrophysics/IMAPP, Radboud University Nijmegen, P.O. Box 9010, 6500 GL 
Nijmegen, The Netherlands }
\altaffiltext {54}{Artemis, Universit\'e C\^ote d'Azur, CNRS, Observatoire C\^ote d'Azur, CS 34229, F-06304 
Nice Cedex 4, France }
\altaffiltext {55}{Institut de Physique de Rennes, CNRS, Universit\'e de Rennes 1, F-35042 Rennes, France }
\altaffiltext {56}{Washington State University, Pullman, WA 99164, USA }
\altaffiltext {57}{Universit\`a degli Studi di Urbino 'Carlo Bo', I-61029 Urbino, Italy }
\altaffiltext {58}{INFN, Sezione di Firenze, I-50019 Sesto Fiorentino, Firenze, Italy }
\altaffiltext {59}{University of Oregon, Eugene, OR 97403, USA }
\altaffiltext {60}{Laboratoire Kastler Brossel, UPMC-Sorbonne Universit\'es, CNRS, ENS-PSL Research 
University, Coll\`ege de France, F-75005 Paris, France }
\altaffiltext {61}{Carleton College, Northfield, MN 55057, USA }
\altaffiltext {62}{Astronomical Observatory Warsaw University, 00-478 Warsaw, Poland }
\altaffiltext {63}{VU University Amsterdam, 1081 HV Amsterdam, The Netherlands }
\altaffiltext {64}{University of Maryland, College Park, MD 20742, USA }
\altaffiltext {65}{Laboratoire des Mat\'eriaux Avanc\'es (LMA), CNRS/IN2P3, F-69622 Villeurbanne, France }
\altaffiltext {66}{Universit\'e Claude Bernard Lyon 1, F-69622 Villeurbanne, France }
\altaffiltext {67}{Universit\`a di Napoli 'Federico II', Complesso Universitario di Monte S.Angelo, I-80126 
Napoli, Italy }
\altaffiltext {68}{NASA/Goddard Space Flight Center, Greenbelt, MD 20771, USA }
\altaffiltext {69}{RESCEU, University of Tokyo, Tokyo, 113-0033, Japan. }
\altaffiltext {70}{University of Adelaide, Adelaide, South Australia 5005, Australia }
\altaffiltext {71}{Tsinghua University, Beijing 100084, China }
\altaffiltext {72}{Texas Tech University, Lubbock, TX 79409, USA }
\altaffiltext {73}{The University of Mississippi, University, MS 38677, USA }
\altaffiltext {74}{The Pennsylvania State University, University Park, PA 16802, USA }
\altaffiltext {75}{National Tsing Hua University, Hsinchu City, 30013 Taiwan, Republic of China }
\altaffiltext {76}{Charles Sturt University, Wagga Wagga, New South Wales 2678, Australia }
\altaffiltext {77}{University of Chicago, Chicago, IL 60637, USA }
\altaffiltext {78}{Kenyon College, Gambier, OH 43022, USA }
\altaffiltext {79}{Korea Institute of Science and Technology Information, Daejeon 305-806, Korea }
\altaffiltext {80}{University of Cambridge, Cambridge CB2 1TN, UK }
\altaffiltext {81}{Universit\`a di Roma 'La Sapienza', I-00185 Roma, Italy }
\altaffiltext {82}{University of Brussels, Brussels 1050, Belgium }
\altaffiltext {83}{Sonoma State University, Rohnert Park, CA 94928, USA }
\altaffiltext {84}{Montana State University, Bozeman, MT 59717, USA }
\altaffiltext {85}{Center for Interdisciplinary Exploration \& Research in Astrophysics (CIERA), Northwestern 
University, Evanston, IL 60208, USA }
\altaffiltext {86}{Universitat de les Illes Balears, IAC3---IEEC, E-07122 Palma de Mallorca, Spain }
\altaffiltext {87}{The University of Texas Rio Grande Valley, Brownsville, TX 78520, USA }
\altaffiltext {88}{Bellevue College, Bellevue, WA 98007, USA }
\altaffiltext {89}{Institute for Plasma Research, Bhat, Gandhinagar 382428, India }
\altaffiltext {90}{The University of Sheffield, Sheffield S10 2TN, UK }
\altaffiltext {91}{California State University, Los Angeles, 5154 State University Dr, Los Angeles, CA 90032, 
USA }
\altaffiltext {92}{Universit\`a di Trento, Dipartimento di Fisica, I-38123 Povo, Trento, Italy }
\altaffiltext {93}{INFN, Trento Institute for Fundamental Physics and Applications, I-38123 Povo, Trento, 
Italy }
\altaffiltext {94}{Cardiff University, Cardiff CF24 3AA, UK }
\altaffiltext {95}{Montclair State University, Montclair, NJ 07043, USA }
\altaffiltext {96}{National Astronomical Observatory of Japan, 2-21-1 Osawa, Mitaka, Tokyo 181-8588, Japan }
\altaffiltext {97}{Canadian Institute for Theoretical Astrophysics, University of Toronto, Toronto, Ontario 
M5S 3H8, Canada }
\altaffiltext {98}{MTA E\"otv\"os University, ``Lendulet'' Astrophysics Research Group, Budapest 1117, 
Hungary }
\altaffiltext {99}{School of Mathematics, University of Edinburgh, Edinburgh EH9 3FD, UK }
\altaffiltext {100}{University and Institute of Advanced Research, Gandhinagar, Gujarat 382007, India }
\altaffiltext {101}{IISER-TVM, CET Campus, Trivandrum Kerala 695016, India }
\altaffiltext {102}{University of Szeged, D\'om t\'er 9, Szeged 6720, Hungary }
\altaffiltext {103}{Embry-Riddle Aeronautical University, Prescott, AZ 86301, USA }
\altaffiltext {104}{Tata Institute of Fundamental Research, Mumbai 400005, India }
\altaffiltext {105}{INAF, Osservatorio Astronomico di Capodimonte, I-80131, Napoli, Italy }
\altaffiltext {106}{University of Michigan, Ann Arbor, MI 48109, USA }
\altaffiltext {107}{Rochester Institute of Technology, Rochester, NY 14623, USA }
\altaffiltext {108}{NCSA, University of Illinois at Urbana-Champaign, Urbana, IL 61801, USA }
\altaffiltext {109}{University of Bia{\l }ystok, 15-424 Bia{\l }ystok, Poland }
\altaffiltext {110}{SUPA, University of Strathclyde, Glasgow G1 1XQ, UK }
\altaffiltext {111}{University of Southampton, Southampton SO17 1BJ, UK }
\altaffiltext {112}{University of Washington Bothell, 18115 Campus Way NE, Bothell, WA 98011, USA }
\altaffiltext {113}{Institute of Applied Physics, Nizhny Novgorod, 603950, Russia }
\altaffiltext {114}{Seoul National University, Seoul 151-742, Korea }
\altaffiltext {115}{Inje University Gimhae, 621-749 South Gyeongsang, Korea }
\altaffiltext {116}{National Institute for Mathematical Sciences, Daejeon 305-390, Korea }
\altaffiltext {117}{Pusan National University, Busan 609-735, Korea }
\altaffiltext {118}{NCBJ, 05-400 \'Swierk-Otwock, Poland }
\altaffiltext {119}{Institute of Mathematics, Polish Academy of Sciences, 00656 Warsaw, Poland }
\altaffiltext {120}{Monash University, Victoria 3800, Australia }
\altaffiltext {121}{Hanyang University, Seoul 133-791, Korea }
\altaffiltext {122}{The Chinese University of Hong Kong, Shatin, NT, Hong Kong }
\altaffiltext {123}{University of Alabama in Huntsville, Huntsville, AL 35899, USA }
\altaffiltext {124}{ESPCI, CNRS, F-75005 Paris, France }
\altaffiltext {125}{University of Minnesota, Minneapolis, MN 55455, USA }
\altaffiltext {126}{Universit\`a di Camerino, Dipartimento di Fisica, I-62032 Camerino, Italy }
\altaffiltext {127}{Southern University and A\&M College, Baton Rouge, LA 70813, USA }
\altaffiltext {128}{The University of Melbourne, Parkville, Victoria 3010, Australia }
\altaffiltext {129}{College of William and Mary, Williamsburg, VA 23187, USA }
\altaffiltext {130}{Instituto de F\'\i sica Te\'orica, University Estadual Paulista/ICTP South American 
Institute for Fundamental Research, S\~ao Paulo SP 01140-070, Brazil }
\altaffiltext {131}{Whitman College, 345 Boyer Avenue, Walla Walla, WA 99362 USA }
\altaffiltext {132}{Universit\'e de Lyon, F-69361 Lyon, France }
\altaffiltext {133}{Hobart and William Smith Colleges, Geneva, NY 14456, USA }
\altaffiltext {134}{Janusz Gil Institute of Astronomy, University of Zielona G\'ora, 65-265 Zielona G\'ora, 
Poland }
\altaffiltext {135}{King's College London, University of London, London WC2R 2LS, UK }
\altaffiltext {136}{IISER-Kolkata, Mohanpur, West Bengal 741252, India }
\altaffiltext {137}{Indian Institute of Technology, Gandhinagar Ahmedabad Gujarat 382424, India }
\altaffiltext {138}{Andrews University, Berrien Springs, MI 49104, USA }
\altaffiltext {139}{Universit\`a di Siena, I-53100 Siena, Italy }
\altaffiltext {140}{Trinity University, San Antonio, TX 78212, USA }
\altaffiltext {141}{University of Washington, Seattle, WA 98195, USA }
\altaffiltext {142}{Abilene Christian University, Abilene, TX 79699, USA }

% pulsar astronomer authors
\author{S.~Buchner\altaffilmark{143,144},  % South Africa
I.~Cognard\altaffilmark{145,146},          % Nancay
A.~Corongiu\altaffilmark{147},             % Cagliari group
P.~C.~C.~Freire\altaffilmark{148},         % Bonn group
L.~Guillemot\altaffilmark{145,146},        % Nancay group
G.~B.~Hobbs\altaffilmark{149},             % ATNF group
M.~Kerr\altaffilmark{149},                 % ATNF group
A.~G.~Lyne\altaffilmark{150},              % Jodrell Bank
A.~Possenti\altaffilmark{147},             % Cagliari group
A.~Ridolfi\altaffilmark{148},              % Bonn group
R.~M.~Shannon\altaffilmark{151,152},       % ATNF group
B.~W.~Stappers\altaffilmark{150},          % Jodrell Bank
and P.~Weltevrede\altaffilmark{150}        % Jodrell Bank
} 

\altaffiltext{143}{Square Kilometer Array South Africa, The Park, Park Road, Pinelands, Cape Town 7405, South 
Africa}
\altaffiltext{144}{Hartebeesthoek Radio Astronomy Observatory, PO Box 443, Krugersdorp, 1740, South 
Africa}
\altaffiltext{145}{Laboratoire de Physique et Chimie de l'Environnement et de l'Espace, LPC2E, 
CNRS-Universit\'{e} d'Orl\'{e}ans, F-45071 Orl\'{e}ans, France}
\altaffiltext{146}{Station de Radioastronomie de Nan\c{c}ay, Observatoire de Paris, CNRS/INSU, F-18330 
Nan\c{c}ay, France}
\altaffiltext{147}{INAF - Osservatorio Astronomico di Cagliari, via della Scienza 5, 09047 Selargius, Italy}
\altaffiltext{148}{Max-Planck-Institut f\"{u}r Radioastronomie MPIfR, Auf dem H\"{u}gel 
69, D-53121 Bonn, Germany}
\altaffiltext{149}{CSIRO Astronomy and Space Science, Australia Telescope National Facility, Box 76 
Epping, NSW, 1710, Australia}
\altaffiltext{150}{Jodrell Bank Centre for Astrophysics, School of Physics and Astronomy, University 
of Manchester, Manchester M13 9PL, UK}
\altaffiltext{151}{CSIRO Astronomy and Space Science, Australia Telescope National Facility, Box 76 
Epping, NSW, 1710, Australia}
\altaffiltext{152}{International Centre for Radio Astronomy Research, Curtin University, Bentley, WA 
6102, Australia}

\fi

\begin{abstract}
We present the result of searches for gravitational waves from \NPULSARS pulsars using data from the first 
observing run of the Advanced LIGO detectors. We find no significant evidence for a \ghw signal from any of 
these pulsars, but we are able to set the most constraining upper limits yet on their \ghw amplitudes and 
ellipticities. For \NBELOWSDWORD of these pulsars, our upper limits give bounds that are improvements over the 
indirect spin-down limit values. For another \WITHINFACTORTEN, we are within a factor of 10 of the 
spin-down limit, and it is likely that some of these will be reachable in future runs of the advanced 
detector. Taken as a whole, these new results improve on previous limits by more than a factor of two.
\end{abstract}

\maketitle

% acronym definitions
\acrodef{MCMC}[MCMC]{Markov chain Monte Carlo}
\acrodef{PWNe}[PWNe]{pulsar-wind nebulae}
\acrodef{TOA}[TOA]{time-of-arrival}

%%%%%%%%%%%%%%%%%

\section{Introduction}\label{sec:intro}

The recent observations of \gws from the inspiral and merger of binary black holes herald the era of \ghw 
astronomy \citep{2016PhRvL.116f1102A, 2016PhRvL.116x1103A}. Such cataclysmic, transient, and extragalatic events 
are not however the only potential sources of observable \gws. Galactic neutron stars offer a more local, 
and continuous, quasi-monochromatic source of gravitational radiation. Although intrinsically far weaker than 
the transient sources that have been observed, their continuous nature allows their signals to be found buried 
deep in the noise by coherently integrating over the long observing runs of the \ghw observatories.

The subset of known pulsars, identified through electromagnetic observations, provides an important possible source of 
continuous \gws. They are often timed with exquisite precision, allowing their rotational phase evolution, 
sky location and, if required, binary orbital parameters to be determined very accurately. In turn, these 
timings allow us to carry out fully phase-coherent and computationally cheap \ghw searches over the length of 
our observation runs. A selection of known pulsars have already been targeted using data from the initial LIGO, 
Virgo, and \geo detectors \citep[summarized in][]{2014ApJ...785..119A}, setting upper limits on their signal 
amplitudes, though without any detections.

An important milestone is passed when this upper limit falls below the so-called spin-down limit on 
gravitational strain for the targeted pulsar. This spin-down limit is determined by equating the power 
radiated through \ghw emission to the pulsar's observed spin-down luminosity (attributed to its loss in rotational 
kinetic energy), i.e.\ as would be the case if it were a {\it gravitar} \citep{Palomba:2005,Knispel:2008}, and determining the 
equivalent strain expected at the Earth.\footnote{This is known to be a na\"{i}ve limit. For several young 
pulsars where the braking index (see Section~\ref{sec:results}) is measured
\citep{2015MNRAS.446..857L,2016ApJ...819L..16A}, we know that it is not consistent with pure \gw emission, and 
other energy-loss mechanisms can be dominant. Effects of this on spin-down limit calculations are discussed 
in \citet{Palomba:2000}. Figures~9 and 10 of \citet{2013ApJS..208...17A} also show that for pulsars 
observed as {\it Fermi} gamma-ray sources, a not insignificant proportion of their spin-down luminosity is 
emitted through gamma-rays.} It can be calculated \citep[see, e.g.][]{2014ApJ...785..119A} using
\begin{equation}\label{eq:h0sd}
h_0^{\rm sd} = \left(\frac{5}{2} \frac{G I_{zz} |\dot{f}_{\rm rot}|}{c^3 d^2 \frot}\right)^{1/2},
\end{equation}
where $f_{\rm rot}$ and $\dot{f}_{\rm rot}$ are the pulsar's frequency and first frequency derivative, 
$I_{zz}$ is the principal moment of inertia (for which we generally assume a canonical value of 
$10^{38}$\,kg\,m$^2$), and $d$ is the pulsar's distance. In previous searches, this limit has been surpassed 
(i.e.\ a smaller limit on the strain amplitude has been obtained) for two pulsars: PSR\,J0534+2200 \citep[the 
Crab pulsar;][]{Abbott:2008} and PSR\,J0835\textminus4510 \citep[the Vela pulsar;][]{Abadie:2011}.

\begin{figure*}[!htbp]
 \includegraphics[width=1.0\textwidth]{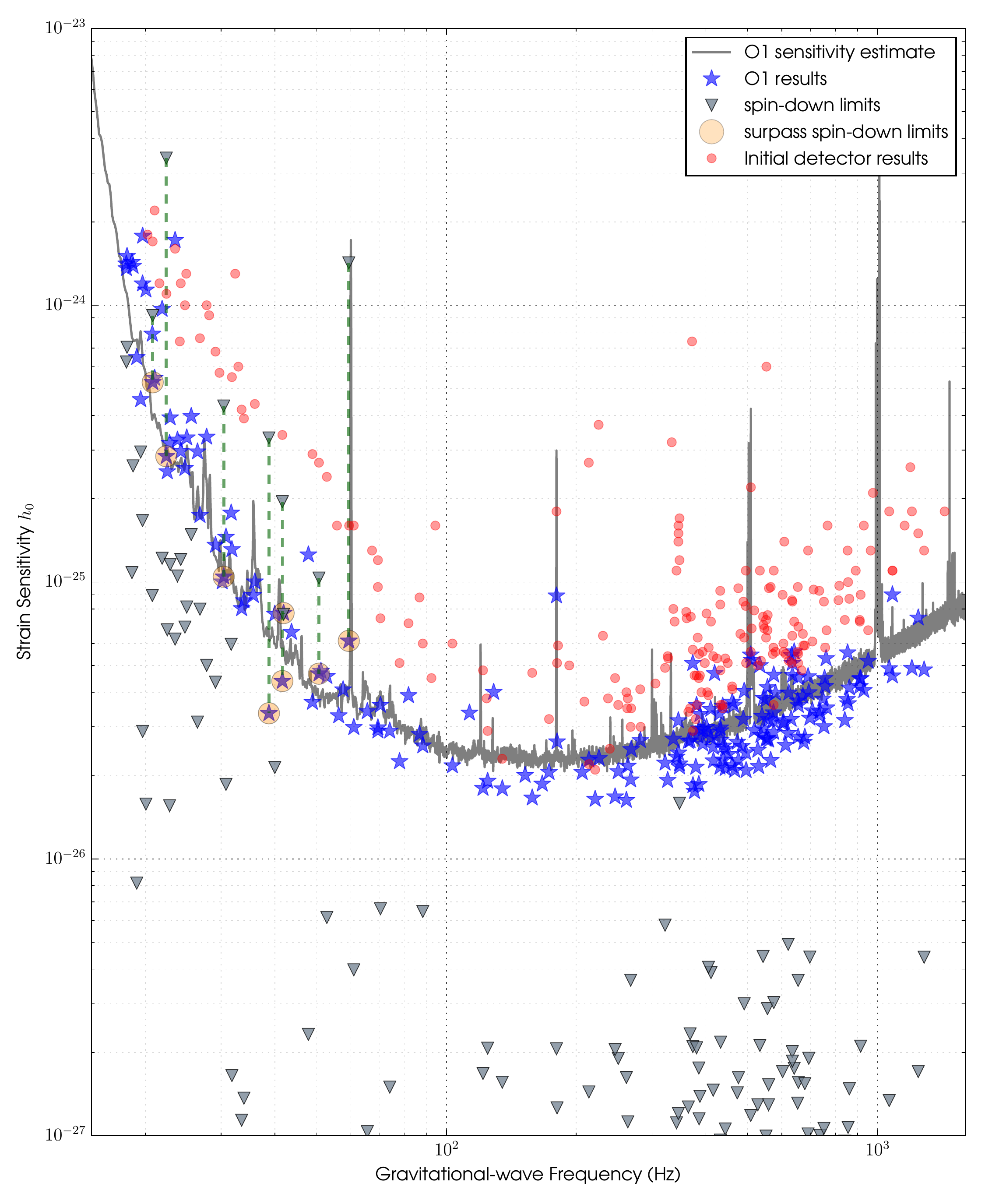}
 \caption{Stars show 95\% credible upper limits on \ghw amplitude, $h_0^{95\%}$, for \NPULSARS pulsars 
using data from the O1 run. $\blacktriangledown$ give the spin-down limits for all pulsars (based on 
distance values taken from the ATNF pulsar catalog \citep{Manchester:2005}, unless otherwise stated in Tables~\ref{tab:highvalue} 
and \ref{tab:allresults}) and assuming the canonical moment of inertia. The upper limits 
shown within the shaded circles are those for which the spin-down limits (linked via the dashed vertical lines) 
are surpassed with our observations. The gray curve gives an estimate of the expected strain sensitivity for 
O1, combining representative amplitude spectral density measurements for both H1 and L1. This estimate is an
angle-averaged value and for particular sources is representative only, whilst the broader range over all
angles for such an estimate is shown, for example, in Figure~4 of \citet{Abbott:2010}. Previous initial
detector run results \citep{2014ApJ...785..119A} for 195 pulsars are shown as red circles, with \NPREVIOUS
of these sources corresponding to sources searched for in O1.
\label{fig:sensest}}
\end{figure*}

In this paper, we provide results from a search for \gws from \NPULSARS known pulsars using data from the first
observing run (O1) of Advanced LIGO (aLIGO). For the LIGO Hanford Observatory (H1) and LIGO Livingston
Observatory (L1), we used data starting on 2015 September 11 at 01:25:03 UTC and 18:29:03 UTC, respectively, and 
finishing on 2016 January 19 at 17:07:59~UTC at both sites. With duty factors of \LHODUTYFACTOR and 
\LLODUTYFACTOR for H1 and L1, this run provided \LHOOBSTIME days and \LLOOBSTIME days of data respectively for analysis.
The estimated sensitivity of this search as a function of source frequency is shown in 
Figure~\ref{fig:sensest}.\footnote{The sensitivity is taken as $10.8\sqrt{S_n'}$, where 
$S_n'$ is the harmonic mean of the observation-time-weighted one-sided power spectral densities, $S_n/T$, for 
H1 and L1 (see \url{https://dcc.ligo.org/LIGO-G1600150/public} and
\url{https://dcc.ligo.org/LIGO-G1600151/public}, respectively). The factor of 10.8 gives the 95\% credible 
upper limit on \ghw strain amplitude averaged over orientation angles assuming Gaussian noise
\citep{Dupuis:2005}.} We see that, even with its comparatively short observation time, the O1 data provide a 
significant sensitivity improvement over the previous runs, particularly at lower frequencies.

\subsection{The signal}

We model the source as a rigidly rotating triaxial star, generating a strain signal at the detector of 
\citep[e.g.][]{1998PhRvD..58f3001J}
\begin{align}\label{eq:signal}
h(t) = &h_0 \bigg[\frac{1}{2}F^D_+(t, \alpha, \delta, \psi)(1+\cos{}^2\iota)\cos{\phi(t)} 
\nonumber \\
& + F^D_{\times}(t,\alpha, \delta, \psi)\cos{\iota}\sin{\phi(t)}\bigg]
\end{align}
where $h_0$ is the \ghw strain amplitude, and $F^D_+$ and $F^D_{\times}$ are the antenna responses of 
observatory $D$ to the `+' and `$\times$' polarizations. These are dependent on the source sky position (right 
ascension $\alpha$ and declination $\delta$) and polarization angle $\psi$.  $\iota$ is the inclination of 
the star's rotation axis to the line of sight, and $\phi(t)$ represents the evolution of the sinusoidal 
signal phase with time.

This phase evolution is usefully represented as a Taylor expansion, so that
\begin{equation}\label{eq:taylor}
\phi(t) = \phi_0 + 2\pi\sum_{j=0}^N
\frac{\mathrel{\overset{\makebox[0pt]{\mbox{\tiny (j)}}}{f_0}}}
{(j+1)!}\left(t-T_0+\delta t(t)\right)^{(j+1)},
\end{equation}
where $\phi_0$ is the initial \ghw phase at time epoch $T_0$, and
$\mathrel{\overset{\makebox[0pt]{\mbox{\tiny (j)}}}{f_0}}$ is the $j^{\rm th}$ time derivative of the \ghw
frequency defined at $T_0$.  $\delta t(t)$ is the time delay from the observatory to the solar system 
barycenter, and can also include binary system barycentering corrections to put the observatory and source in 
inertial frames. For the majority of pulsars, expansions to $N=1$ or 2 are all that are required, but for 
some young sources, with significant timing noise, expansions to higher orders may be used. For the case of a
source rotating around a principal axis of inertia and producing emission from the $l=m=2$ (spherical harmonic)
mass quadrupole mode (e.g.\ a rigidly rotating star with a triaxial moment of inertia ellipsoid), the \ghw
frequencies and frequency derivatives are all twice their rotational values, e.g.\ $f = 2\frot$.

\section{Pulsar selection}\label{sec:pulsars}

To reflect the improved sensitivity of LIGO during O1, we targeted pulsars with rotation frequencies, \frot, 
of greater than about 10\,Hz, but also included seven promising sources with large spin-down 
luminosities\footnote{PSRs J0908\tmin4913, J1418\tmin6058, J1709\tmin4429, J1826\tmin1334, J1845\tmin0743, 
J1853\tmin0004, and J2129+1210A} with \frot just below 10\,Hz. The $l=m=2$ quadrupolar emission frequencies 
of these targets are therefore greater than $\sim20$\,Hz and within the band of good sensitivity for the 
instruments. We did not impose an upper limit on target frequency.

We have obtained timings for \NPULSARS known pulsars in this band. Timing was performed using the 42\,ft
telescope and Lovell telescope at Jodrell Bank (UK), the 26\,m telescope at Hartebeesthoek (South 
Africa), the Parkes radio telescope (Australia), the Nan\c{c}ay Decimetric Radio Telescope (France), the 
Arecibo Observatory (Puerto Rico) and the {\it Fermi} Large Area Telescope (LAT). Of these, \NPREVIOUS of these have 
been targeted in previous campaigns \citep{2014ApJ...785..119A}, whilst \NNEW are new to this search.

For the vast majority of these, we have obtained timing solutions using pulse \ac{TOA} observations that
spanned the O1 run. For those pulsars whose \acp{TOA} did not span O1, we still expect them to maintain very 
good coherence when extrapolated to the O1 time. The {\sc tempo}\footnote{\url{http://tempo.sourceforge.net}} 
or {\sc tempo2} \citep{Hobbs:2006} pulsar timing codes were used to produce these solutions, which provide us 
with precise information on the parameters defining each pulsars phase evolution, including their sky 
location and any binary system orbital dynamics if applicable.\footnote{Of the 200 pulsars, 119 are in binary systems.}

\subsection{High-value targets}\label{sec:highvalue}

We identified \NHIGHVALUE sources (Table~\ref{tab:highvalue}) for which we could either improve upon, or 
closely approach, the spin-down limit based on Equation~\ref{eq:h0sd}. These are all young pulsars at the 
lower end of our sensitive frequency band and include the Crab and Vela pulsars for which the spin-down limit 
had already been surpassed \citep{Abbott:2008, Abadie:2011, 2014ApJ...785..119A}.

\begin{deluxetable}{l c c c}
\tablecaption{The high-value targets for which the spin-down limit can be improved upon or closely
approached.\label{tab:highvalue}}
\tablehead{
\colhead{PSR} &
\colhead{$f$ (Hz)} &
\colhead{$d$ (kpc)} &
\colhead{$h_0^{\rm spin-down}$}}
\startdata
J0205+6449\tablenotemark{$\dagger$} & 30.4 & 3.2\phantom{$^{*\#}$} & $4.3\ee{-25}$ \\
J0534+2200 (Crab) & 59.3 & 2.0\phantom{$^{*\#}$} & $1.4\ee{-24}$ \\
J0835\textminus4510 (Vela) & 22.4 & 0.3\phantom{$^{*\#}$} & $3.4\ee{-24}$ \\
J1302\textminus6350\tablenotemark{$\ddagger$} & 41.9 & 2.3\phantom{$^{*\#}$} & $7.7\ee{-26}$ \\
J1809\textminus1917 & 24.2 & 3.7\phantom{$^{*\#}$} & $1.2\ee{-25}$ \\
J1813\textminus1246 & 41.6 & 2.5\tablenotemark{$*$}\phantom{$^{\#}$} & $2.0\ee{-25}$ \\
J1826\textminus1256 & 18.1 & 1.2\tablenotemark{$\#$}\phantom{$^{*}$} & $7.1\ee{-25}$ \\
J1928+1746 & 29.1 & 8.1\phantom{$^{*\#}$} &  $4.4\ee{-26}$ \\
J1952+3252 (CTB\,80) & 50.6 & 3.0\phantom{$^{*\#}$} & $1.0\ee{-25}$ \\
J2043+2740 & 20.8 & 1.1\phantom{$^{*\#}$} & $9.2\ee{-25}$ \\
J2229+6114 & 38.7 & 3.0\phantom{$^{*\#}$} & $3.3\ee{-25}$
\enddata
\tablenotetext{$\dagger$}{This pulsar was observed to glitch during O1 on MJD 57345.}
\tablenotetext{$\ddagger$}{This pulsar is in a binary system and as such was not able to be searched for
with the $5n$-vector method.}
\tablenotetext{$*$}{This distance is a lower limit on the distance from \citet{2014ApJ...795..168M}. It is
slightly higher than the distance of 1.9\,kpc used for calculations in \citet{2014ApJ...785..119A}.}
\tablenotetext{$\#$}{This distance is that taken from the lower distance range from
\citet{2016MNRAS.458.2813V} \citep[using values from][]{Wang:2011}.}
\tablecomments{Unless otherwise stated, all distances are those from {\tt v1.54} of the ATNF Pulsar Catalog 
\citep{Manchester:2005}.}
\end{deluxetable}

\section{Analyses}\label{sec:analyses}

Following  \citet{2014ApJ...785..119A}, we used three largely independent methods for carrying out the search
for the \NHIGHVALUE high-value targets: the time-domain-based {\it Bayesian}
\citep{Dupuis:2005} and $\mathcal{F}$/$\mathcal{G}$-{\it statistic} \citep{Jaranowski:2010} methods, and the
frequency-domain-based {\it 5$n$-vector} method \citep{2010CQGra..27s4016A,Astone:2012}. For the other 
\NLOWVALUE targets only the {\it Bayesian} method was applied.

We refer the reader to \citet{2014ApJ...785..119A} and references therein for more detailed descriptions of
these methods. Generally, the methods were not modified for O1, although there have been some significant 
improvements to the {\it Bayesian} method, which are described in Appendix~\ref{app:bayesian}.

In addition, the results from the $5n$-vector method used an earlier data release, with a slightly different 
instrumental calibration~\citep{2016arXiv160203845T}, than that used for the two other methods. The calibrations applied differ, however, 
by less than 3\% in amplitude and less than $3^{\circ}$ in phase for all high-value sources.

For one high-value target, PSR\,J1302\textminus6350, the {\it 5$n$-vector} method was not used. This pulsar is in a
binary system, which is not currently handled by this method. PSR\,J0205+6449 underwent a glitch on MJD 57345 
(2015 November 19), causing the rotation frequency to increase by $\sim 8.3\ee{-6}$\,Hz. Because of the uncertain 
relation between the \ghw  and electromagnetic signal phases over a glitch, we analyzed both the 
pre-and-post-glitch periods independently and combined these incoherently to give the final result. To the 
best of our knowledge, none of our other sources glitched during the course of O1.

The results from the {\it Bayesian} method incorporate uncertainties into the pulsars' phase evolutions. If
the fits to pulsar \acp{TOA} from electromagnetic observations provided uncertainties on any fitted
parameters, then these parameters were also included in the search space (in addition to the four main 
unknown signal parameters, $h_0$, $\phi_0$, $\cos{\iota}$ and $\psi$, defined with equations~\ref{eq:signal} and \ref{eq:taylor}). Prior probabilities 
for these additional parameters were defined as Gaussian distributions, using their best-fit values and 
associated errors as means and standard deviations \citep[see, e.g.][]{Abbott:2010}. Upper limits are produced 
from the posterior probability distributions on $h_0$, by marginalizing all other parameters over their prior 
ranges (see Appendix~\ref{app:prior}) and calculating the $h_0$ value bounding (from zero) 95\% of the 
probability \citep[e.g., Equation~3.3 of][]{Abbott:2007a}.

Observations of \ac{PWNe} around several pulsars allow us to put prior constraints on their orientation angles 
$\iota$ and $\psi$, detailed in Appendix~\ref{app:restricted}. For these pulsars, any results given include 
both those based on the standard prior ranges for the orientation angles given in Equation~\ref{eq:priors}, 
as well as those based on these restricted ranges.

\section{Results}\label{sec:results}

For all pulsars, we quote 95\% credible/confidence upper limits on the \ghw amplitude $h_0$ set using 
coherently combined data from both H1 and L1.\footnote{For the Bayesian results, these are credible limits 
bounded from zero, whilst for the frequentist results these are confidence limits.} We use this value to also 
set limits on the mass quadrupole moment $Q_{22}$ of the $l=m=2$ mode of the star \citep{Owen:2005} via
\begin{equation}\label{eq:q22}
 Q_{22} = h_0\left( \frac{c^4 d}{16\pi^2G \frot^2} \right) \sqrt{\frac{15}
{8\pi}}.
\end{equation}
In turn, this is related to the star's fiducial equatorial ellipticity 
$\varepsilon$ through 
\begin{equation}\label{eq:epsilon}
\varepsilon = \frac{Q_{22}}{I_{zz}} \sqrt{\frac{8\pi}{15}}.
\end{equation}
To calculate $\varepsilon$, we use the canonical moment of inertia of $I_{zz} = 10^{38}$\,kg\,m$^2$ \citep[see,
e.g., Chapter 6 of][]{PulsarAstronomy}. We also 
quote the ratio of our observed $h_0$ limits to the spin-down limits calculated using Equation~\ref{eq:h0sd}. 
The distances used to calculate $Q_{22}$ and $\varepsilon$ are (unless otherwise stated in 
Table~\ref{tab:highvalue} or \ref{tab:allresults}) taken from {\tt v1.54} of the ATNF pulsar catalog
\citep{Manchester:2005},\footnote{\url{http://www.atnf.csiro.au/people/pulsar/psrcat/}} and in most cases 
are calculated from the observed dispersion measure \citep[noting that distance uncertainties of 20\% or more 
are not uncommon; see, e.g.\ Figure~12 of][]{2002astro.ph..7156C}. For the spin-down limit calculation, we 
generally use values of $\dot{f}_{\rm rot}$ provided from the electromagnetic-pulse-arrival-time fits used in 
our search. If, however, an intrinsic period derivative, i.e.\ a period derivative corrected for proper motion 
effects \citep{Shklovskii:1969} or globular cluster accelerations, is given in the ATNF catalog, then that 
value is used. If an intrinsic period derivative is not given for a globular cluster pulsar, then the spin-down 
limit is instead based on an assumed characteristic spin-down age of $\tau = 10^9$\,yr. The characteristic 
age \citep[see, e.g., Chapter 6 of][]{PulsarAstronomy} is defined as
\begin{equation}\label{eq:tau}
\tau = -\frac{f_{\rm rot}}{\dot{f}_{\rm rot}(n-1)},
\end{equation}
where $n$ is the braking index ($n=f_{\rm rot}\ddot{f}_{\rm rot}/\dot{f}_{\rm rot}^2$), which has a value of 
$n=3$ for purely magnetic dipole radiation, whilst we adopt the $n=5$ case for purely gravitational radiation.

The calibration procedure for the aLIGO instruments and their amplitude uncertainties during the initial part 
of O1 are described in detail in \citet{2016arXiv160203845T}. After O1 was completed, the calibration was updated, 
and the maximum calibration uncertainties estimated over the whole run give a $1\sigma$ limit on the combined
H1 and L1 amplitude uncertainties of $\lesssim 14\%$. This is the conservative level of uncertainty on the
$h_0$ upper limits, and any quantities derived linearly from them, from the \ghw observations alone.

\begin{figure}[!htbp]
 \includegraphics[width=0.49\textwidth]{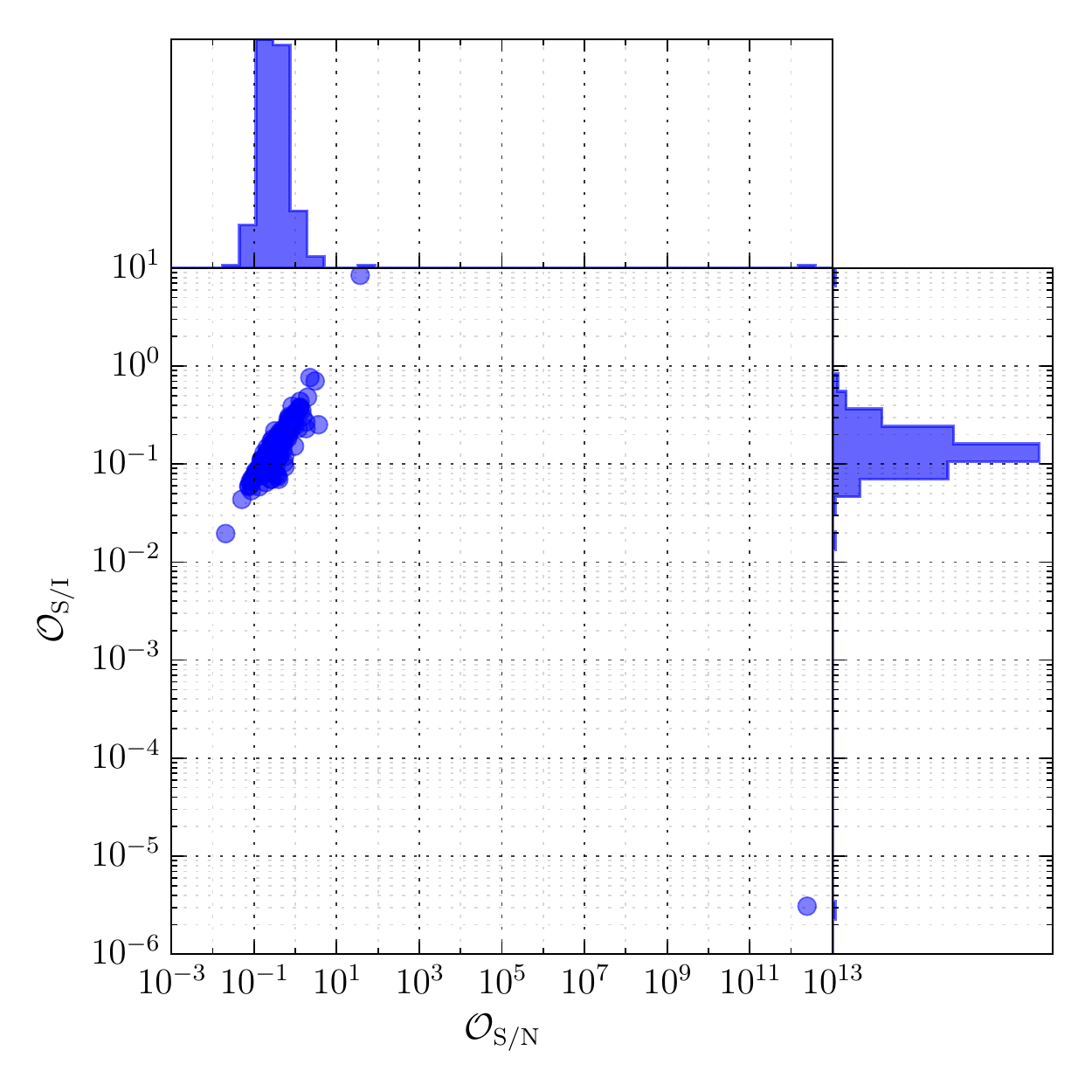}
 \caption{Distributions of the probability ratios $\mathcal{O}_{\textrm{S/N}}$ and $\mathcal{O}_{\rm
{S/I}}$ for the observed pulsars.
\label{fig:odds}}
\end{figure}

The results for all targets, except the high-value targets discussed in Section~\ref{sec:highvalue}, are shown 
in Table~\ref{tab:allresults}. For each pulsar, we produce two probability ratios, or odds (discussed in 
Appendix~\ref{app:evidence}): $\mathcal{O}_{\textrm{S/N}}$, Equation~\ref{eq:or1}, comparing the probability 
that the data from both detectors contain a coherent signal matching our model to the probability that they 
both contain just (potentially non-stationary) Gaussian noise; and, $\mathcal{O}_{\rm {S/I}}$, 
Equation~\ref{eq:oddsratio}, comparing the probability that the data from both detectors contain a coherent 
signal matching our model to the probability of the data containing combinations of independent signals or 
noise. The latter of these is an attempt to account for incoherent interference in the detectors (e.g.\ 
produced by instrumental line artifacts) that can mimic the effects of a signal. The distributions of these 
odds for all our sources can be seen in Figure~\ref{fig:odds}.\footnote{For each source, a different prior 
volume was used, so directly comparing odds values between sources should be treated with caution.} We find 
that the largest ratio for $\mathcal{O}_{\rm {S/I}}$ is 8 for PSR\,J1932+17. Although this is larger than any other 
source and favors a coherent signal over the alternative incoherent-\textit{or}-noise hypothesis by over a 
factor of eight, it is not yet strong enough evidence for a signal \citep[e.g.\ in the interpretation scaling 
of][]{Jeffreys:1931}, especially considering the multiple searches that are performed. The largest 
$\mathcal{O}_{\textrm{S/N}}$ value is for PSR\,J1833\tmin0827, with a value of $2.5\ee{12}$ in favor of the 
signal model. However, as is apparent from the $\mathcal{O}_{\rm {S/I}}$ value of $3\ee{-6}$ and the 
posterior distributions of parameters, it is clear that the very large $\mathcal{O}_{\rm S/N}$ comes from 
strong interference in the data, whilst there is no support for a coherent signal in both detectors.

The $h_0$ upper limits from this analysis (including those from the high-value targets) are shown in
Figure~\ref{fig:sensest}. The figure also contains the upper limits obtained for the 195 pulsars targeted
using data from the initial detector era \citep{2014ApJ...785..119A}. We find that, on average, for pulsars 
that were both analyzed here and in previous runs, our new results have over two and a half times better 
sensitivity. The largest improvement is a factor of eight for PSR\,J0024\tmin7204C at $f=347.4$\,Hz. For four 
pulsars, the new results are slightly less sensitive than the previous analyses, although in the worst 
case this is only by $\lesssim 10\%$.\footnote{This is for PSR\,J1833\textminus0827 at $23.4$\,Hz, for which 
there appears to be a large amount of incoherent interference between the detectors.}

\begin{figure*}[!htbp]
 \includegraphics[width=1.0\textwidth]{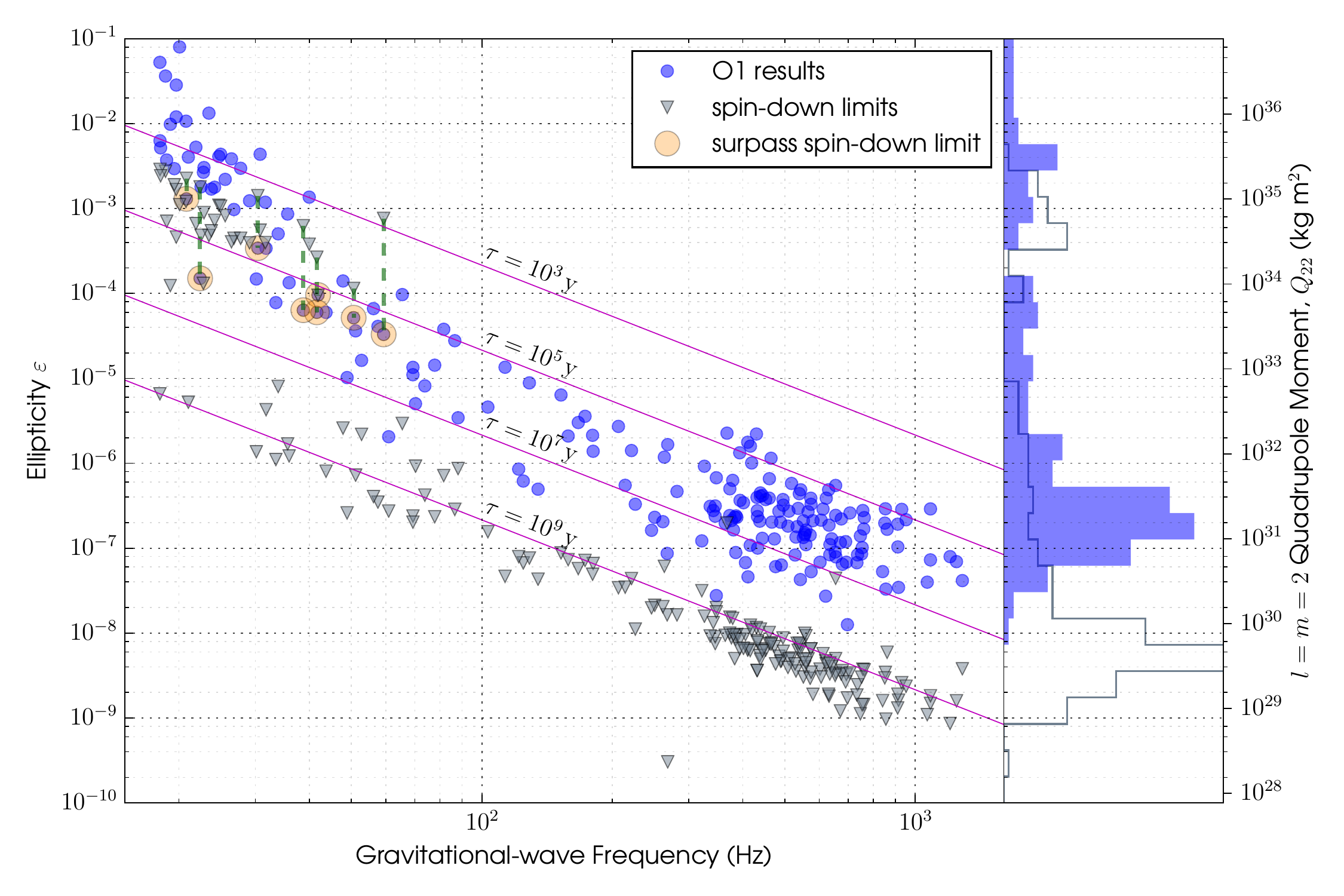}
 \caption{Limits on fiducial ellipticities ($\varepsilon$) and mass quadrupole moments ($Q_{22}$).
$\blacktriangledown$ show the values based on the spin-down limits for these pulsars. The pulsars for which
the spin-down limit is surpassed are highlighted within larger shaded circles and linked to their spin-down 
limit values with dashed vertical lines. Also shown are diagonal lines of constant characteristic age, 
$\tau$, for gravitars (with braking indices of $n=5$) calculated via $\varepsilon^{\rm sd} = 1.91\ee{5} 
f_{\rm rot}^{-2}/\sqrt{(n-1)\tau I_{38}}$, where $I_{38}$ is the principal moment of inertia in units of 
$10^{38}$\,kg\,m$^2$ (where we set $I_{38}=1$). \label{fig:ellq22}}
\end{figure*}

\begin{figure}[!htbp]
 \includegraphics[width=0.49\textwidth]{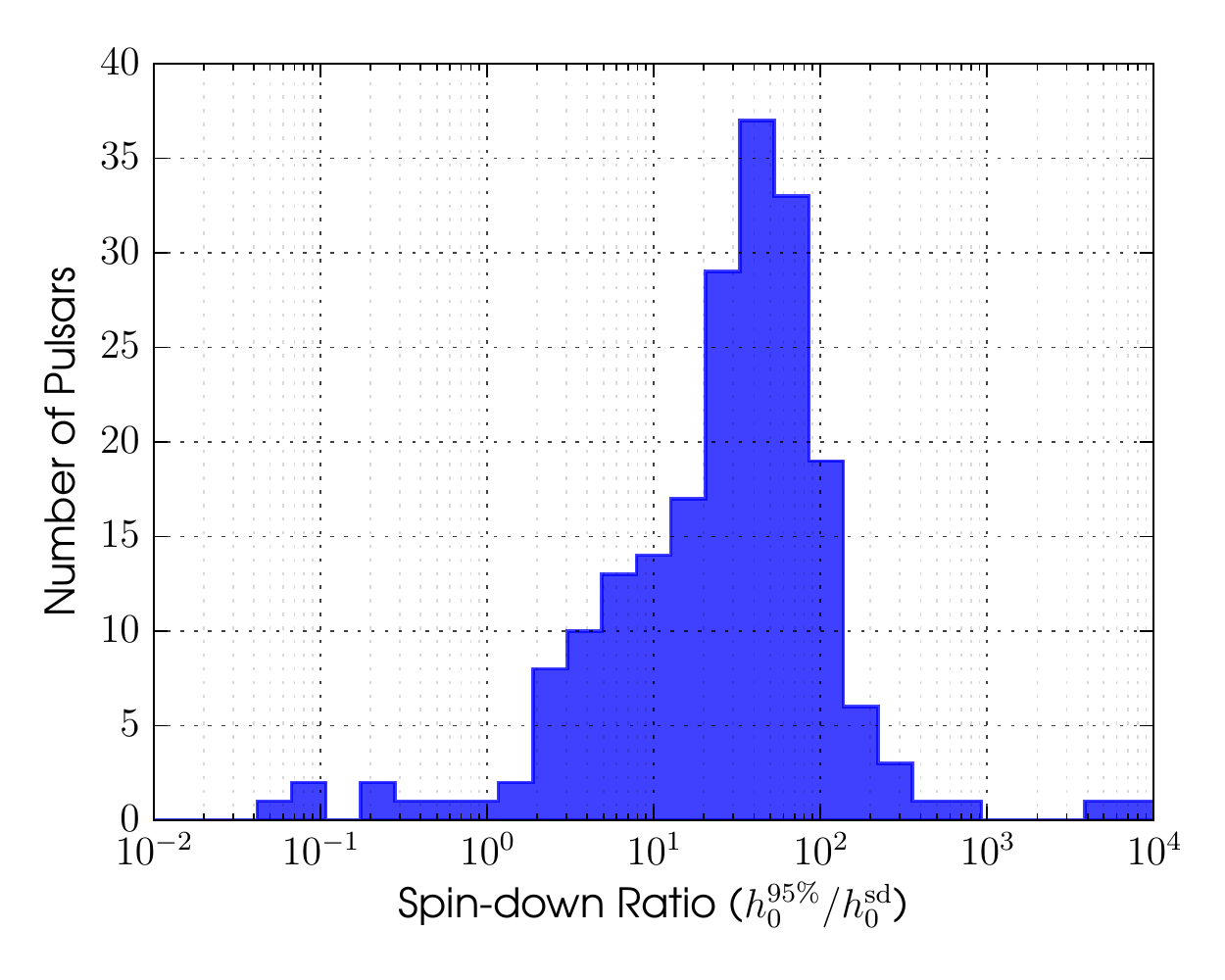}
 \caption{Ratio between our observed $h_0^{95\%}$ limits and the spin-down limits for all pulsars.
\label{fig:ratios}}
\end{figure}

Figure~\ref{fig:ellq22} shows corresponding limits on the fiducial ellipticity $\varepsilon$ and mass
quadrupole moment $Q_{22}$. Figure~\ref{fig:ratios} shows a histogram of the ratios between our upper limits 
and the spin-down limits.

The accelerations that pulsars experience in the cores of globular clusters can mask their true spin-down 
values. It is sometimes possible to determine these accelerations and correct for their effect on spin-down.
As mentioned above, when such a correction is available, we have calculated the spin-down limits based on this
corrected spin-down value. In cases where the correction is not available we have instead assumed each pulsar
has a characteristic age of $\tau = 10^9$\,yr and under the assumption of gravitational-radiation-dominated
spin-down, calculated a na\"{i}ve spin-down via Equation~\ref{eq:tau}, which has then been used for
the spin-down limit calculation. As proposed in \citet{Pitkin:2011}, for these pulsars we could instead
invert the process and use the $h_0$ upper limit to set a limit on the spin-down of the pulsars (at least
under the assumption that they are gravitars, with $n=5$). Given that the maximum observed spin-up for a
globular cluster pulsar is $\sim 5\ee{-14}$\,Hz\,s$^{-1}$, we can say that the negative of this can be used as 
an approximation for the largest magnitude {\it spin-down} that could be masked by intracluster
accelerations.\footnote{In \citet{2006CQGra..23S...1O} and \citet{Pitkin:2011}, it is stated that
$-5\ee{-13}$\,Hz\,s$^{-1}$ is roughly the largest magnitude spin-down that could be masked by globular cluster
accelerations. This is mainly based on the maximum observed spin-up for a globular cluster pulsar
(PSR\,J2129+1210D) being $\sim 5\ee{-13}$\,Hz\,s$^{-1}$ as given in {\tt v1.54} of the ATNF pulsar catalog
\citep{Manchester:2005}. However, this value appears to be wrong, with the original observations for
PSR\,J2129+1210D \citep{Anderson:1993} giving a value of just under $\sim 5\ee{-14}$\,Hz\,s$^{-1}$. This is
still the maximum observed spin-up for any globular cluster pulsar.} Of the globular cluster pulsars for
which the intrinsic spin-down is not known, we find that our upper limits on $h_0$ give the smallest limit on
the absolute spin-down value due to \gws for PSR\,J1623\tmin2631 of $\dot{f} = -3.2\ee{-13}$\,Hz\,s$^{-1}$.
Although this value is probably too large to be masked by accelerations, it is of the same order as the 
spin-downs for two globular cluster millisecond pulsars, PSRs J1823-3021A \citep{2011Sci...334.1107F} and
J1824-2452A \citep{2013ApJ...778..106J}, both with apparently large intrinsic spin-down values.

\subsection{High-value targets}

Table~\ref{tab:highvalueres} shows the results for the high-value targets (Section~\ref{sec:highvalue}) for
each of the three analysis methods discussed in Section~\ref{sec:analyses}. The results from the methods are 
broadly consistent. For pulsars that have restricted priors on orientations, the results using these are shown 
alongside the results from the full prior orientation range. We find that for \NBELOWSDWORD of these pulsars, 
we achieve a sensitivity that surpasses the indirect spin-down limit.

Table~\ref{tab:highvalueres} also contains an estimate of the maximum surface deformation of the $l=m=2$ 
mode, $R\varepsilon_{\rm surf,22}$, for each of the pulsars. This is based on Figure~2 of
\citet{Johnson-McDaniel:2013}, where we adopt a scaling of $R\varepsilon_{\rm surf,22} \approx
25(\varepsilon/10^{-4})$\,cm maximized over equations of state and possible stellar masses. We also find
that for five of these pulsars (PSRs J0534+2200, J1302\tmin6350, J1813\tmin1246, J1952+3252, and J2229+6114) 
the $l=m=2$ surface deformations are smaller than the rotational ($l=2$, $m=0$) surface deformation for all 
equations of state.\footnote{For this we have assumed a 1.4\,M$_{\odot}$ star and used approximate scalings 
calculated from Table~1 of \citet{Johnson-McDaniel:2013}, taking into account that the rotational deformation 
scales with $f_{\rm rot}^2$.} For the Vela pulsar (PSR\,J0835\tmin4510) and PSR\,J0205+6449, the $l=m=2$ 
surface deformations are smaller than the rotational deformations for roughly half of the equations of state 
used in \citet{Johnson-McDaniel:2013}. There is no expected relation between the scales of these two 
deformations, but it is intriguing to compare them nonetheless.

\section{Discussion}\label{sec:discussion}

We have searched for \ghw emission from the $l=m=2$ quadrupole mode of \NPULSARS known pulsars. There is no
significant evidence for a signal from any of the sources. We have been able to set 95\% credible upper
limits on the \ghw amplitudes from all these sources, and from these derived limits on each star's
fiducial ellipticity and quadrupole moment.

In earlier analyses, the indirect spin-down limits on the \ghw amplitude had been surpassed for two pulsars:
PSR\,J0534+2200 \citep[the Crab pulsar;][]{Abbott:2008} and PSR\,J0835\textminus4510 \citep[the Vela 
pulsar;][]{Abadie:2011}. We improve upon the previous limits for these two pulsars by factors of $\gtrsim 
3$. We find that for the Crab and Vela pulsars, less than $\sim 2\ee{-3}$ and $\sim 10^{-2}$ of the 
spin-down luminosity is being lost via gravitational radiation, respectively (assuming the distance is precisely 
known and using the fiducial moment of inertia of $10^{38}$\,kg\,m$^2$). The observed braking indices of these pulsars provide
constraints on the contribution of \ghw emission to the spin-down, under the assumption that the spin-down is due only to a
combination of electromagnetic and \ghw losses. These braking index constraints are more stringent, i.e.\ give smaller limits on
the \ghw emission, than the na\"ive spin-down limit given in Equation~\ref{eq:h0sd} \citep[see][]{Palomba:2000}. Our results, however,
surpass even these more stringent limits and are therefore compatible with the observed braking indices. We surpass 
the spin-down limits of six further pulsars. All these are young pulsars with large spin-down luminosities, 
and as such our limits translate to large ellipticities/quadrupole moments that are at the upper end of some 
maximally allowed values \citep[see e.g.][]{Owen:2005,Pitkin:2011,Johnson-McDaniel:2012}. If we assume that 
internal toroidal magnetic fields are the source of any stellar mass quadrupole \citep{Bonazzola:1996}, then 
we can use our limits on ellipticities as constraints on the magnitude of the internal field strength. For
the Crab pulsar, PSR\,J1813\tmin1246, PSR\,J1952+3252, and PSR\,J2229+6114, which have roughly comparable 
ellipticity limits, the internal magnetic field strength is limited to $\lesssim 10^{16}$\,G 
\citep[e.g.][]{Cutler:2002, Haskell:2007bh}. For comparison, the Crab pulsar's inferred external polar 
magnetic field at its surface is $\sim 4\ee{12}$\,G. Due to this being a rough order of magnitude estimate, 
this value is the same as that previously quoted for the Crab pulsar in \citet{2014ApJ...785..119A}, although 
the limit is now valid for several more pulsars.

For any neutron star equation of state, the lower bound on the mass quadrupole (due to the internal magnetic field,
which may be very weak) is many orders of magnitude less than the upper bound.  Therefore, it is always important
to acknowledge that these upper limits on particular stars do not allow us to place constraints on neutron star
equations of state.

Of all the pulsars, the smallest 95\% credible limit on $h_0$ that we find is \MINHZERO for \MINHZEROPULSAR.
The smallest ellipticity and $Q_{22}$ quadrupole moments are \MINELL and \MINQ, respectively, for \MINEQPULSAR,
which is a relatively nearby pulsar at \MINEQDIST. Although neither of these pulsars surpasses their
fiducial spin-down limits, it is interesting to note that there are \WITHINFACTORTEN that we are able to
constrain to within a factor of 10 of their spin-down limits (see Figure~\ref{fig:ratios}). For 
PSR\,J0437\textminus4715 (which is nearby, at 0.16\,kpc), we are in fact only 1.4 times above the spin-down 
limit. Therefore, an equivalent increase in detector sensitivity of that factor, or a $1.4^2\approx 1.9$ 
times longer run, would allow us to surpass the spin-down limit. Alternatively, the spin-down limit would be 
surpassed if the true moment of inertia for PSR\,J0437\textminus4715 were a factor of 1.9 times larger 
than $I_{38}$, which is well within plausible values. As this is a millisecond pulsar, it would give 
ellipticity constraints of less than a few $10^{-8}$, or $l=m=2$ quadrupole moment constraints of 
$\lesssim 10^{30}$\,kg\,m$^2$, compared to the much larger constraints typically found for the young pulsars 
in Table~\ref{tab:highvalue}. Using the conversion in \citet{Cutler:2002} the constraints on the internal 
toroidal fields for this pulsar would be $\lesssim 10^{13}$\,G, which is similar to the external field 
strengths of young pulsars.

This search has imposed a model in which the \ghw signal phase evolutions must be tightly locked to the
pulsars' rotational evolutions determined through electromagnetic observations. There are mechanisms
\citep[discussed in, e.g.][]{Abbott:2008}, however, that could lead to small deviations between the phase
evolution and observed rotation. Additionally, there are many pulsars for which highly accurate
timings do not exist or are not available from observations coincident with ours.\footnote{One desirable
source that we no longer have accurate timings for is PSR\,J0537\textminus6910, an X-ray pulsar in the Large
Magellanic Cloud, for which we relied on the now-defunct RXTE satellite.} There are several such sources for
which the spin-down limit could be surpassed and these are being searched for in O1 data using narrow-band
searches \citep[see, e.g.][]{narrowb:2015}, covering a small range in frequency and frequency derivative to
account for uncertainties in the exact parameters (LIGO Scientific Collaboration \& Virgo Collaboration 2017, in
preparation). All-sky broadband searches for unknown rotating neutron stars are also underway.

In the near future, increasing sensitivities and considerably longer observing runs are planned for aLIGO and
Advanced Virgo \citep{2016LRR....19....1A}. This will give us several times greater sensitivity with which to
search for \ghw signals, and in any event will allow us to surpass the spin-down limits for 10 or 
more pulsars. Future searches will also address \ghw emission at not just twice the rotation frequency, but 
also at the rotation frequency \citep[e.g.][]{2015MNRAS.453.4399P}, further increasing the likelihood of a 
first detection of continuous gravitational waves.

\acknowledgements

% September 2015 LVC acknowledgements https://dcc.ligo.org/LIGO-M1500315
The authors gratefully acknowledge the support of the United States National Science Foundation (NSF) for the 
construction and operation of the LIGO Laboratory and Advanced LIGO as well as the Science and Technology 
Facilities Council (STFC) of the United Kingdom, the Max-Planck-Society (MPS), and the State of
Niedersachsen/Germany for support of the construction of Advanced LIGO and construction and operation of the 
GEO600 detector. Additional support for Advanced LIGO was provided by the Australian Research Council. The 
authors gratefully acknowledge the Italian Istituto Nazionale di Fisica Nucleare (INFN), the French Centre 
National de la Recherche Scientifique (CNRS), and the Foundation for Fundamental Research on Matter supported 
by the Netherlands Organisation for Scientific Research, for the construction and operation of the Virgo 
detector and the creation and support of the EGO consortium. The authors also gratefully acknowledge research 
support from these agencies as well as by the Council of Scientific and Industrial Research of India, 
Department of Science and Technology, India, Science \& Engineering Research Board (SERB), India,
Ministry of Human Resource Development, India, the Spanish Ministerio de Econom\'ia y Competitividad,
the Vicepresid\`encia i Conselleria d'Innovaci\'o, Recerca i Turisme and the Conselleria d'Educaci\'o i
Universitat of the Govern de les Illes Balears, the National Science Centre of Poland, the European 
Commission, the Royal Society, the 
Scottish Funding Council, the Scottish Universities Physics Alliance, the Hungarian Scientific Research Fund 
(OTKA), the Lyon Institute of Origins (LIO), the National Research Foundation of Korea, Industry Canada and 
the Province of Ontario through the Ministry of Economic Development and Innovation, the Natural Science and 
Engineering Research Council Canada, Canadian Institute for Advanced Research, the Brazilian Ministry of 
Science, Technology, and Innovation, Funda\c{c}\~ao de Amparo \`a Pesquisa do Estado de S\~ao Paulo (FAPESP),
Russian Foundation for Basic Research, the Leverhulme Trust, the Research Corporation, Ministry of Science 
and Technology (MOST), Taiwan and the Kavli Foundation. The authors gratefully acknowledge the support of the 
NSF, STFC, MPS, INFN, CNRS and the State of Niedersachsen/Germany for provision of computational resources.

Pulsar observations with the Lovell telescope and their analyses are supported through a consolidated grant 
from the STFC in the UK. The Nan\c{c}ay Radio Observatory is operated by the Paris Observatory, associated 
with the French CNRS. A.~Ridolfi and P.~C.~C.~Freire gratefully acknowledge financial support by the European 
Research Council for the ERC Starting grant BEACON under contract no.\ 279702. A.~Ridolfi is member of the 
International Max Planck research school for Astronomy and Astrophysics at the Universities of Bonn and 
Cologne and acknowledges partial support through the Bonn-Cologne Graduate School of Physics and Astronomy.

LIGO Document No.\ LIGO-P1600159.

\appendix

\section{The application of the Bayesian method}\label{app:bayesian}

The {\it Bayesian method} used in our known pulsar searches involves a data processing stage and a parameter estimation stage. For a given
source, the data processing stage takes calibrated strain data from H1 and L1 (sampled at a rate of
16\,384\,Hz), heterodynes it to remove a best fit for the source's phase evolution, and then low-pass filters
and heavily down-samples the data to one sample per minute \citep{Dupuis:2005}. This leaves a complex time
series with a $1/60$\,Hz bandwidth, in which a signal would have the form
\begin{equation}\label{eq:h2f}
h'(t) = e^{i\Delta\phi(t)}\frac{h_0}{2}\bigg(\frac{1}{2}F_{+}(\psi,t)[1+\cos{}^2\iota]e^{i\phi_{0}}
-iF_{\times}(\psi,t)\cos{\iota}e^{i\phi_{0}} \bigg),
\end{equation}
where the remaining modulation will be due to the detector's diurnal antenna pattern and any slowly varying
phase difference caused by potential differences between the best-fit phase evolution and the true signal 
phase evolution $\Delta \phi(t) = \left(\phi_{\rm true}(t)-\phi_{\rm best-fit}(t)\right)$.\footnote{The 
analysis code actually works with a signal parameterized in terms of the ``waveform'' model defined in 
\citet{2015MNRAS.453...53J} and \citet{2015MNRAS.453.4399P}, where $h_0 = -2C_{22}$ and $\phi_0 = \Phi_{22}^C$.}

These combined processed datasets for each detector $\mathbf{d}$ are used to estimate the joint posterior
probability distribution of the unknown source signal parameters, $\vec{\theta}$, using a Bayesian framework
via
\begin{equation}\label{eq:bayes}
p(\vec{\theta}|\mathbf{d},H_{\mathrm{S}},I) = \frac{p(\mathbf{d}|\vec{\theta},H_{\mathrm{S}},I)
p(\vec{\theta}|H_{\mathrm{S}},I)}{p(\mathbf{d}|H_{\mathrm{S}},I)},
\end{equation}
where $p(\mathbf{d}|\vec{\theta},H_{\mathrm{S}},I)$ is the likelihood of the data given the specific signal
model ($H_{\textrm{S}}$) parameters, $p(\vec{\theta}|H_{\mathrm{S}},I)$ is the joint prior probability
distribution of the parameters, and $p(\mathbf{d}|H_{\mathrm{S}},I)$ is the evidence (marginal likelihood) of
observing our data, given a signal of the type we defined. In the cases where $\Delta \phi(t)$ are negligible,
this corresponds to just estimating four parameters, \ $\vec{\theta} = \left\{h_0, \phi_0, \cos{\iota},
\psi\right\}$. In general, offsets between the best-fit phase parameters and true signal parameters can also
be estimated, provided that they do not cause the signal to drift out of the bandwidth available. When using
timing solutions calculated using {\sc tempo(2)}, uncertainties in the fitted parameters are produced, and when
available these fitted parameters will be included in our estimation for the \ghw signal.

In previous searches \citep[e.g.~][]{Abbott:2010, 2014ApJ...785..119A}, a \ac{MCMC} method has been used to
sample and estimate $p(\vec{\theta}|\mathbf{d},H_{\mathrm{S}},I)$ for the unknown parameters. However, the
simple proposal distribution used for the \ac{MCMC} was not well-tuned and was therefore inefficient, 
especially when searching over additional phase parameters. Furthermore, the \ac{MCMC} did not naturally
produce a value for the evidence $p(\mathbf{d}|H_{\mathrm{S}},I)$. To allow the calculation of
$p(\mathbf{d}|H_{\mathrm{S}},I)$ and, as a natural by-product, the joint parameter posterior probability
distribution, we have adopted the nested sampling method \citep{Skilling:2006}. In particular our analysis
code (\citet{2012JPhCS.363a2041P}; M.~Pitkin et al.\ 2017, in preparation) uses the nested sampling implementation of \citet{Veitch:2010}
as provided in the {\tt LALIinference} library \citep{2015PhRvD..91d2003V} within the LIGO Algorithm Library
(LAL) suite.\footnote{\url{https://wiki.ligo.org/DASWG/LALSuite}} This implements more intelligent and 
efficient proposals than previously used.  The code has been validated by extracting both software and 
hardware \citep{Biwer:2016} signal injections into \ghw detector data.

\LongTables
\begin{deluxetable*}{l c c c c c c c c c c c c c}
\tabletypesize{\scriptsize}
\tablecaption{Limits on the \ghw amplitude, and other derived quantities, for the \NHIGHVALUE high-value
pulsars\label{tab:highvalueres}}

\tablehead{
\colhead{Analysis} &
\colhead{$h_0^{95\%}$} &
\colhead{$\varepsilon$} &
\colhead{$Q_{22}$} &
\colhead{$h_0^{95\%}/h_0^{\rm sd}$} &
\colhead{$\dot{E}_{\rm gw}/\dot{E}$} &
\colhead{$R\varepsilon_{\rm surf,22}$\,cm\tablenotemark{$\ddagger$}} &
\colhead{$\log_{10} \mathcal{O}_{\textrm{S/I}}$\tablenotemark{$\sharp$}/} \\
\colhead{} &
\colhead{($10^{-25}$)} &
\colhead{($10^{-4}$)} &
\colhead{($10^{34}$\,kg\,m$^2$)} &
\colhead{} &
\colhead{} &
\colhead{} &
\colhead{FAP\tablenotemark{$\spadesuit$}/$p$-value\tablenotemark{$\dagger$}}
}

\startdata
\cutinhead{PSR\,J0205+6449 (Pre-glitch)}

Bayesian & 1.1 (1.3) & 3.6 (4.3) & 2.8 (3.3) & 0.25 (0.31) & 0.064 (0.093) & 90 (110) & $-1.2$ ($-1.1$) \\

\FG-statistic & 1.8 (2.4) & 5.9 (7.9) & 4.5 (6.1) & 0.42 (0.55) & 0.17 (0.31) & 150 (200) & 0.27 (0.16) \\

$5n$-vector & 0.75 (1.1) & 2.5 (3.7) & 1.9 (2.8) & 0.17 (0.25) & 0.030 (0.064) & 60 
(90) & 0.95 \\

\cutinhead{(Post-glitch)}

Bayesian & 2.0 (2.6) & 6.4 (8.4) & 5.0 (6.5) & 0.45 (0.60) & 0.21 (0.36) & 160 (210) & $-1.0$ ($-0.7$) \\

\FG-statistic & 1.7 (1.4) & 5.6 (4.6) & 4.3 (3.5) & 0.39 (0.32) & 0.15 (0.10) & 140 (120) & 0.49 (0.91) \\

$5n$-vector & 1.1 (1.7) & 3.6 (5.4) & 2.8 (4.2) & 0.25 (0.38) & 0.065 (0.15) & 90 (140) 
& 0.32 \\

\cutinhead{(Incoherently Combined)}

Bayesian & 1.0 (1.3) & 3.4 (4.4) & 2.6 (3.4) & 0.24 (0.31) & 0.058 (0.097) & 90 (110) & $-0.6$ ($-0.5$) \\

\FG-statistic & 1.2 (1.6) & 3.9 (5.2) & 3.0 (4.0) & 0.28 (0.37) & 0.077 (0.14) & 100 (130) & 0.36 (0.48) \\

$5n$-vector & 0.73 (1.1) & 2.3 (3.5) & 1.8 (2.7) & 0.17 (0.25) & 0.028 (0.064) & 60 
(90) & 0.95 \\

\cutinhead{PSR\,J0534+2200 (Crab)}

Bayesian & 0.67 (0.61) & 0.36 (0.33) & 0.28 (0.25) & 0.05 (0.04) & 0.0022 (0.0018) & 9 (8) & $-0.7$ ($-0.7$) 
\\

\FG-statistic & 0.42 (0.24) & 0.23 (0.13) & 0.17 (0.10) & 0.03 (0.02) & 0.00087 (0.00028) & 6 (3) & 0.62 (0.31) \\

$5n$-vector & 0.52 (0.50) & 0.28 (0.27) & 0.22 (0.21) & 0.04 (0.04) & 0.0013 (0.0012) & 7 (7) & 0.21 \\

\cutinhead{PSR\,J0835\textminus4510 (Vela)}

Bayesian & 3.2 (2.8) & 1.7 (1.5) & 1.3 (1.2) & 0.10 (0.08) & 0.0090 (0.0070) & 40 (40) & $-0.9$ ($-0.9$) \\

\FG-statistic & 3.8 (3.3) & 2.0 (1.7) & 1.6 (1.3) & 0.11 (0.10) & 0.012 (0.0094) & 50 (40) & 0.37 (0.58) \\

$5n$-vector & 2.9 (2.9) & 1.5 (1.5) & 1.2 (1.2) & 0.09 (0.09) & 0.0073 (0.0073) & 40 (40) & 0.66 \\

\cutinhead{PSR\,J1302\textminus6350}

Bayesian & 0.77 & 0.96 & 0.74 & 1.0 & 1.0 & 20 & $-1.0$ \\

\F-statistic & 0.60 & 0.74 & 0.58 & 0.78 & 0.61 & 20 & 0.44  \\

\cutinhead{PSR\,J1809\textminus1917}

Bayesian & 3.0 & 18 & 14 & 2.5 & \nodata & 450 & $-1.0$ \\

\F-statistic & 2.4 & 14 & 11 & 2.0 & \nodata & 350 & 0.72  \\

$5n$-vector & 2.5 & 15 & 12 & 2.1 & \nodata & 380 & 0.62 \\

\cutinhead{PSR\,J1813\textminus1246}

Bayesian & 0.44 & 0.60 & 0.46 & 0.23 & 0.051 & 20 & $-1.2$ \\

\F-statistic & 0.55 & 0.75 & 0.58 & 0.28 & 0.079 & 20 & 0.61 \\

$5n$-vector & 0.55 & 0.75 & 0.58 & 0.28 & 0.079 & 20 & 0.69 \\

\cutinhead{PSR\,J1826\textminus1256}

Bayesian & 15 & 52 & 40 & 2.1 & \nodata & 1300 & $-0.9$ \\

\F-statistic & 17 & 59 & 45 & 2.4 & \nodata & 1500 & 0.29 \\

$5n$-vector & 18 & 62 & 48 & 2.6 & \nodata & 1600 & 0.21 \\

\cutinhead{PSR\,J1928+1746}

Bayesian & 1.4 & 12 & 9.5 & 3.1 & \nodata & 300 & $-1.0$ \\

\F-statistic & 1.5 & 14 & 11 & 3.4 & \nodata & 350 & 0.42 \\

$5n$-vector & 1.3 & 12 & 9.1 & 3.0 & \nodata & 300 & 0.70 \\

\cutinhead{PSR\,J1952+3252}

Bayesian & 0.47 (0.50) & 0.52 (0.56) & 0.40 (0.43) & 0.45 (0.49) & 0.20 (0.24) & 10 (10) & $-1.1$ ($-1.1$) \\

\F-statistic & 0.48 & 0.53 & 0.41 & 0.46 & 0.22 & 10 & 0.44 \\

$5n$-vector & 0.37 (0.39) & 0.41 (0.43) & 0.32 (0.33) & 0.36 (0.38) & 0.13 (0.14) & 10 (10) & 0.57 \\

\cutinhead{PSR\,J2043+2740}

Bayesian & 5.3 & 13 & 10 & 0.57 & 0.32 & 330 & $-0.8$ \\

\F-statistic & 5.6 & 14 & 11 & 0.61 & 0.37 & 350 & 0.41 \\

$5n$-vector & 6.0 & 15 & 11 & 0.65 & 0.43 & 380 & 0.18 \\

\cutinhead{PSR\,J2229+6114}

Bayesian & 0.50 (0.34) & 0.95 (0.64) & 0.73 (0.49) & 0.15 (0.10) & 0.023 (0.010) & 20 (20) & $-1.3$ ($-1.4$) 
\\

\FG-statistic & 0.49 (0.45) & 0.93 (0.85) & 0.72 (0.66) & 0.15 (0.14) & 0.022 (0.018) & 20 (20) & 0.73 (0.35)
\\

$5n$-vector & 0.56 (0.43) & 1.1 (0.84) & 0.82 (0.63) & 0.17 (0.13) & 0.029 (0.017) & 30 
(20) & 0.59 \\

\enddata
\tablecomments{Limits with constrained orientations (see Appendix~\ref{app:restricted}) are given in 
parentheses. When the spin-down limit is not surpassed, no power ratio, $\dot{E}_{\rm gw}/\dot{E}$, is given.}

\tablenotetext{$\ddagger$}{This is the equivalent upper limit on the $l=m=2$ surface deformation maximized 
over the equation of state and stellar mass \citep{Johnson-McDaniel:2013}. Values below 10 are rounded to the 
nearest integer, values between 10 and 1000 are rounded to the nearest decade, and values above 1000 are 
rounded to the nearest hundred.}

\tablenotetext{$\sharp$}{For the Bayesian analysis, this column gives the logarithm of the odds for a
coherent signals being present in the data versus an incoherent signal or noise being present in the data
(equation~\ref{eq:oddsratio}).}

\tablenotetext{$\spadesuit$}{For the \F/\G-statistic analysis this column gives the false alarm probability.
The false alarm probabilities are calculated using the observed values of 2\F and 2\G, and assuming they
are drawn from $\chi^2$ distributions with 4 and 2 degrees of freedom for the \F- and \G-statistics, respectively.}

\tablenotetext{$\dagger$}{For the $5n$-vector results, this column gives the significance expressed as a 
$p$-value representing the probability that noise alone can produce a value of the detection statistic larger 
than that actually obtained in the analysis \citep[see][for more discussion of this]{narrowb:2015}.}

\end{deluxetable*}

\subsection{The likelihood}\label{app:likelihood}

The likelihood, $p(\mathbf{d}|\vec{\theta},H_{\mathrm{S}},I)$, is a Student's $t$-like probability
distribution and is given in, e.g., \citet{Abbott:2007a}. It assumes that the noise in the data may be
non-stationary, but consists of stationary Gaussian segments, each with unknown variance. The analysis uses a
{\it Bayesian Blocks}-type method \citep{1998ApJ...504..405S} to divide the data into stationary segments,
although those containing fewer than five points are discarded. Any segments longer than a day (1440 points
given our 1/60\,Hz sample rate) are split such that no segments are longer than 1440 points. This differs from
previous analyses in which the data were automatically split into segments containing 30 points.

In cases where the search requires the recalculation of $\Delta \phi(t)$ when evaluating the likelihood,
this can be computationally expensive; the phase, including solar system and binary system barycentring
time delays, is needed and the log-likelihood calculation requires summations over all data points. To make 
this considerably more efficient, we have adopted a {\it Reduced Order Quadrature} scheme
\citep[e.g.][]{Antil2013,2013PhRvD..87l4005C} to approximate the likelihood via interpolation of a reduced
model basis.

\subsection{The priors}\label{app:prior}

In Equation~\ref{eq:bayes}, a prior probability distribution for the parameters is required. For the
parameters $\phi_0$, $\cos{\iota}$, and $\psi$, we generally have no prior knowledge of their values, and so use
flat priors within their allowed ranges:
\begin{align}\label{eq:priors}
p(\phi_0|H_{\mathrm{S}},I) =& \begin{cases} 1/2\pi & \text{if } 0 \leq \phi_0 \leq 2\pi, \\
                             0 & \text{otherwise},
                            \end{cases} \nonumber \\
p(\cos{\iota}|H_{\mathrm{S}},I) =& \begin{cases} 1/2 & \text{if } -1 \leq \cos{\iota} \leq 1, \\
                             0 & \text{otherwise},
                            \end{cases} \\
p(\psi|H_{\mathrm{S}},I) =& \begin{cases} 2/\pi & \text{if } 0 \leq \psi \leq \pi/2, \\
                             0 & \text{otherwise}.
                            \end{cases} \nonumber
\end{align}
These ranges do not necessarily span the full physically allowable range of source values, but are a
degenerate range that will contain all possible observable signal waveforms \citep{2015MNRAS.453...53J,
2015MNRAS.453.4399P}. In some cases, there is information about the inclination and/or polarization angle of the
source (see Appendix~\ref{app:restricted}). Where present this can be incorporated into the prior by using a
Gaussian distribution based on this information. For the cases where the inclination is recovered from a
\ac{PWNe} image, there is no information about the rotation direction of the source, so in fact a bimodal
Gaussian prior on $\iota$ is required \citep{2015MNRAS.453...53J} (see
Appendix~\ref{app:restricted}).\footnote{A bimodal prior was {\it not} used in \citet{2014ApJ...785..119A}, 
but subsequently its inclusion was found to have minimal effect on the upper limits produced.}

For a prior on the \ghw amplitude $h_0$, the analysis in \citet{2014ApJ...785..119A} used a flat
distribution bounded at zero and some value that was large compared to the observed standard deviation of
the data, or a distribution on $h_0$ and $\cos{\iota}$ based on previous searches
\citep[e.g.][]{2010ApJ...713..671A}. In this analysis, inspired by that used in \citet{2016MNRAS.455L..72M},
we have adopted a different prior based on the Fermi-Dirac distribution:
\begin{equation}\label{eq:fermidirac}
p(h_0|\sigma, \mu, H_{\mathrm{S}},I) = \frac{1}{\sigma\log{\left(1+e^{\mu/\sigma}
\right)}}\left(e^{(h_0-\mu)/\sigma} + 1\right)^{-1},
\end{equation}
where $\mu$ gives the point at which the distribution falls to half its maximum value, and $\sigma$ defines
the rate at which the distribution falls off. If we define a value $u^{95\%}$ at which the cumulative
distribution function of Equation~\ref{eq:fermidirac} is at 95\%, and require that the probability density
function falls from 97.5\% to 2.5\% of its maximum over a range that is $0.4\mu$, we are able to define $\mu$
and $\sigma$. In this analysis, there are two ways in which we define $u^{95\%}$ to calculate $\mu$ and
$\sigma$: for pulsars where we already have a 95\% $h_0$ upper limit from previous searches, we use this value
as $u^{95\%}$; for new pulsars, we have based $u^{95\%}$ on the 95\% upper limit that would have been expected
if the pulsar had been searched for in the previous S6/VSR2,4 analysis.\footnote{For two pulsars, PSR\,J0024\textminus7204X
and PSR\,J0721\textminus2038, the priors set using an estimated 95\% upper limit from the S6/VSR2,4 analysis
were found to be too narrow and unduly narrowed the posterior. So, to maintain a more conservative upper limit
dominated by the likelihood, as has been the case in previous searches, the priors were widened by a factor of
three.} For small values of $h_0$ this prior looks flat, whilst for large values it approximates an exponential
distribution. Unlike the flat priors used previously, it is continuous for positive values and penalizes excessively
large values.

If searching over the phase parameters defining $\Delta \phi(t)$ in Equation~\ref{eq:h2f}, i.e.\ frequency,
sky position, and binary system parameters, the prior distribution on the parameters is based on the
uncertainties provided by the {\sc tempo(2)} fits to \acp{TOA}. We take the uncertainties as the standard
deviations for a multivariate Gaussian prior on these parameters. We conservatively have the parameters as
uncorrelated, except in two specific cases for low eccentricity ($e < 10^{-3}$) binary systems. If there are
uncertainties on the time and angle of periastron, or if there are uncertainties on the binary period and time
derivative of the angle of periastron, then these pairs of parameters are set to be fully correlated.

\subsection{The evidence}\label{app:evidence}

The evidence allows a Bayesian model comparison to be performed, i.e.\ the comparison of the relative
probabilities of different signal models given the data, which provides a way of assessing if an observed
signal is real \citep[see, e.g.\ the $\mathcal{B}$-statistic of][for the use of a Bayesian model comparison in this
context]{2009CQGra..26t4013P}. For example, we can calculate the ratio of the probability that the data
contains a signal to the probability that the data is purely Gaussian noise:
\begin{equation}\label{eq:or1}
\mathcal{O}_{\textrm{S/N}} = \frac{p(H_{\mathrm{S}}|\mathbf{d},I)}{p(H_{\mathrm{N}}|\mathbf{d},I)} =
\frac{p(\mathbf{d}|H_{\mathrm{S}},I)}{p(\mathbf{d}|H_{\mathrm{N}},I)}
\frac{p(H_{\mathrm{S}}|I)}{p(H_{\mathrm{N}}|I)}
\end{equation}
where the first term on the right-hand side is called the Bayes factor and
$p(H_{\mathrm{S}}|I)/p(H_{\mathrm{N}}|I)$ is the prior odds of the two models, which we set to be unity. To
calculate $p(\mathbf{d}|H_{\mathrm{N}},I)$ the likelihood can be evaluated with the signal set to zero.

Given more than one detector, we are also able to compare the probability that the data contain a coherent
signal between detectors (as would be expected from an astrophysical source) versus independent (and
therefore incoherent) signals in each detector {\it or} the data consisting of non-stationary (see
Appendix~\ref{app:likelihood}) Gaussian noise alone \citep[e.g.][]{2014PhRvD..89f4023K}. If we take the
combined data to be $\mathbf{d} = \{\mathbf{d}_{\rm H1}, \mathbf{d}_{\rm L1}\}$, then we can form four
incoherent-signal-or-noise hypotheses (where for compactness we have removed the implicit $I$ dependence):
\begin{description}
 \item [$H_{\rm{N}_1}$] an independent signal in both detectors, $p({\mathbf d}|H_{\rm{N}_1}) =
p(\mathbf{d}_{\rm H1}|H_{\mathrm{S}_{\mathrm{H1}}}) p(\mathbf{d}_{\rm L1}|H_{\mathrm{S}_{\mathrm{L1}}})$;
 \item [$H_{\rm{N}_2}$] a signal in H1, but non-stationary Gaussian noise in L1, $p({\mathbf
d}|H_{\rm{N}_2}) =
p(\mathbf{d}_{\rm H1}|H_{\mathrm{S}_{\mathrm{H1}}}) p(\mathbf{d}_{\rm L1}|H_{\mathrm{N}_{\mathrm{L1}}})$;
 \item [$H_{\rm{N}_3}$] a signal in L1, but non-stationary Gaussian noise in H1, $p({\mathbf
d}|H_{\rm{N}_3}) = p(\mathbf{d}_{\rm H1}|H_{\mathrm{N}_{\mathrm{H1}}}) p(\mathbf{d}_{\rm
L1}|H_{\mathrm{S}_{\mathrm{L1}}})$;
 \item [$H_{\rm{N}_4}$] independent non-stationary Gaussian noise in both detectors, $p({\mathbf
d}|H_{\rm{N}_4}) = p(\mathbf{d}_{\rm H1}|H_{\mathrm{N}_{\mathrm{H1}}}) p(\mathbf{d}_{\rm
L1}|H_{\mathrm{N}_{\mathrm{L1}}})$,
\end{description}
where $H_{\rm{S/N}_{\rm{H1/L1}}}$ represents the hypothesis of our signal model/noise in the given detector. 
This gives a ratio
\begin{equation}\label{eq:oddsratio}
 \mathcal{O}_{\textrm{S/I}} = \frac{p(\mathbf{d}|H_{\mathrm{S}})p(H_{\mathrm{S}})}
{p(\mathbf{d}|H_{\mathrm{N}_1}) p(H_{\mathrm{N}_1}) +
p(\mathbf{d}|H_{\mathrm{N}_2}) p(H_{\mathrm{N}_2}) +
p(\mathbf{d}|H_{\mathrm{N}_3}) p(H_{\mathrm{N}_3}) +
p(\mathbf{d}|H_{\mathrm{N}_4}) p(H_{\mathrm{N}_4})}.
\end{equation}
We choose the five hypothesis priors ($p(H_{\mathrm{S}})$, $p(H_{\mathrm{N}_1})$, $p(H_{\mathrm{N}_2})$,
$p(H_{\mathrm{N}_3})$ and $p(H_{\mathrm{N}_3})$) such that they have equal probabilities, and they therefore 
factorize out of the calculation.\footnote{The hypothesis $H_{\rm{N}_1}$ contains all the other hypotheses as 
its subsets, but within it the other hypotheses will all be downweighted by their tiny prior volumes in 
comparison to the full volume. Therefore, to provide more weight to the alternative noise hypotheses, we 
explicitly include them with equal weight.} Such a probability ratio (i.e.\ the odds) obviously
penalizes single detector detections, in which one detector may be considerably more sensitive than the other.

\section{Orientation angle priors}\label{app:restricted}

For several pulsars in our search, there are observations of their \ac{PWNe}. Under the 
assumption that a pulsar's orientation is aligned with its surrounding nebula, we can use the fits to the 
pulsar orientation given in \citet{Ng:2004, Ng:2008} as {\it restricted} priors on $\psi$ and $\iota$. For 
the {\it Bayesian} and {\it 5$n$-vector} methods, the prior probability distributions on $\psi$ and $\iota$ 
are Gaussian distributions based on the \ac{PWNe} fits, whilst the $\mathcal{G}$-{\it statistic} uses a
$\delta$-function prior at the best-fit value. Table~\ref{tab:restricted} shows the means and standard
deviations used for the parameter priors. In general, these are taken from Table~2 of \citet{Ng:2008} where
$\Psi$ is equivalent to our $\psi$ and $\zeta$ is equivalent to our $\iota$.\footnote{$\psi$ can be rotated by 
integer numbers of $\pi/2$~radians and still give signals within our search parameter space, although for 
each rotation any signal would have $\phi_0$ equivalently rotated by $\pi$~radians.} Statistical and
systematic uncertainties are added in quadrature (for non-symmetric uncertainties, the larger value is used). 
For the Crab pulsar and PSR\,J0205+6449 \citet{Ng:2008} give fits to the inner and outer \ac{PWNe} torii, so in 
these cases our mean value is the average of the inner and outer fits, and the quadrature-combined systematic 
and statistical errors for each are combined via $\sigma =
\sqrt{(\sigma_{\rm inner}/2)^2 + (\sigma_{\rm outer}/2)^2}$.

When these restricted priors were used in the previous analyses of \citet{2014ApJ...785..119A},
there has been an implicit (and at the time unrealized) assumption about the rotation of the star.
As noted in \citet{2015MNRAS.453...53J}, constraining $\iota$ and $\psi$ to particular values implicitly
forces a rotation direction on the signal, whilst the \ac{PWNe} observations (or indeed the electromagnetic
timing observations) give us no knowledge of the actual rotation direction. To incorporate this unknown
rotation direction in the search, whilst maintaining the convenient minimal range in $\psi$ of
$\pi/2$~radians, there must be a bimodal distribution on $\iota$ with the additional mode at $\pi -
\iota$~radians. The mean and standard deviations of Gaussian prior distributions used for $\psi$ and the two
modes for $\iota$ are given in Table~\ref{tab:restricted}.

\begin{deluxetable*}{l c c c}
\tabletypesize{\footnotesize}
\tablecaption{Means and standard deviations for restricted priors on
$\psi$ and $\iota$ based on Table~2 of \citet{Ng:2008}.\label{tab:restricted}}
\tablehead{
\colhead{PSR} &
\colhead{$\psi$ (rad)} &
\colhead{$\iota_1$ (rad)} &
\colhead{$\iota_2$ (rad)}}
\startdata
J0205+6449 & \phantom{\tmin}$1.5760 \pm 0.0078$ & $1.5896 \pm 0.0219$ & $1.5519 \pm 0.0219$ \\
J0534+2200 (Crab) & \phantom{\tmin}$2.1844 \pm 0.0016$ &  $1.0850 \pm 0.0149$ & $2.0566 \pm 0.0149$ \\
J0835\textminus4510 (Vela) & \phantom{\tmin}$2.2799 \pm 0.0015$ & $1.1048 \pm 0.0105$ & $2.0368 \pm 0.0105$ \\
J1709\textminus4429 (B1706\textminus44) & \phantom{\tmin}$2.8554 \pm 0.0305$ & $0.9303 \pm 0.0578$ & $2.2113
\pm
0.0578$ \\
J1952+3252 & $-0.2007 \pm 0.1501$ & \nodata & \nodata \\
J2229+6114 & \phantom{\tmin}$1.7977 \pm 0.0454$ & $0.8029 \pm 0.1100$ & $2.3387 \pm 0.1100$
\enddata
\tablecomments{For PSR\,J1952+3252, the values for $\psi$ are not from \ac{PWNe} fitting but are from the
mean of a value derived from proper motion measurements and observations of H$\alpha$ ``lobes''
bracketing the bow shock \citep{Ng:2004}.}
\end{deluxetable*}

\LongTables % table can span multiple pages

\begin{deluxetable*}{l c c c c c c c c c}
\tabletypesize{\footnotesize}
\tablecaption{Limits on the \ghw amplitude, and other derived quantities, for known 
pulsars \label{tab:allresults}}\tablehead{\colhead{PSR} & 
\colhead{$f$ (Hz)} & \colhead{$d$\,(kpc)} & 
\colhead{$h_0^{\rm sd}$} & \colhead{$h_0^{95\%}$} & \colhead{$\varepsilon$} & \colhead{$Q_{22}$} & 
\colhead{Spin-down Ratio} & \colhead{$\log{}_{10}\left(\mathcal{O}_{\textrm{S/N}}\right)$} & 
\colhead{$\log{}_{10}\left(\mathcal{O}_{\textrm{S/I}}\right)$}\\ ~ & ~ & ~ & $(10^{-25})$ & 
$(10^{{-25}})$ 
& $(10^{-7})$ & $(10^{31}$\,kg\,m$^2$) & ~ & ~ & ~}
\startdata
J0023+0923 &655.69 &1.0 &0.016 &0.36 &0.79 &0.61 &23 &$-0.5$ &$-1.0$ \\
J0024\textminus7204AA &1083.79 &4.0 &0.0057\tablenotemark{$\ddagger$} &0.90 &2.9 &2.2 &160 &$0.0$ &$-0.5$ \\
J0024\textminus7204AB &539.86 &4.0 &0.0057\tablenotemark{$\ddagger$} &0.33 &4.4 &3.4 &58 &$-0.4$ &$-0.9$ \\
J0024\textminus7204C &347.42 &4.0 &0.0057\tablenotemark{$\ddagger$} &0.22 &6.8 &5.2 &38 &$-1.1$ &$-1.2$ \\
J0024\textminus7204D &373.30 &4.0 &0.0057\tablenotemark{$\ddagger$} &0.18 &5.0 &3.9 &32 &$-0.5$ &$-0.8$ \\
J0024\textminus7204E &565.56 &4.0 &0.0042\tablenotemark{$\dagger$} &0.23 &2.7 &2.1 &54 &$-0.9$ &$-1.0$ \\
J0024\textminus7204F &762.32 &4.0 &0.0057\tablenotemark{$\ddagger$} &0.35 &2.3 &1.8 &62 &$-0.5$ &$-0.8$ \\
J0024\textminus7204G &495.00 &4.0 &0.0057\tablenotemark{$\ddagger$} &0.24 &3.7 &2.9 &43 &$-0.7$ &$-0.9$ \\
J0024\textminus7204H &622.99 &4.0 &0.0044\tablenotemark{$\dagger$} &0.40 &3.9 &3.0 &90 &$-0.1$ &$-0.6$ \\
J0024\textminus7204I &573.89 &4.0 &0.0057\tablenotemark{$\ddagger$} &0.34 &3.9 &3.0 &60 &$-0.4$ &$-0.7$ \\
J0024\textminus7204J &952.09 &4.0 &0.0057\tablenotemark{$\ddagger$} &0.52 &2.2 &1.7 &91 &$-0.3$ &$-0.7$ \\
J0024\textminus7204L &460.18 &4.0 &0.0057\tablenotemark{$\ddagger$} &0.22 &3.9 &3.0 &38 &$-0.5$ &$-0.8$ \\
J0024\textminus7204M &543.97 &4.0 &0.0057\tablenotemark{$\ddagger$} &0.38 &4.9 &3.8 &67 &$-0.3$ &$-0.7$ \\
J0024\textminus7204N &654.89 &4.0 &0.0057\tablenotemark{$\ddagger$} &0.28 &2.4 &1.9 &49 &$-0.5$ &$-0.8$ \\
J0024\textminus7204O &756.62 &4.0 &0.0057\tablenotemark{$\ddagger$} &0.42 &2.8 &2.1 &74 &$-0.3$ &$-0.8$ \\
J0024\textminus7204Q &495.89 &4.0 &0.0040\tablenotemark{$\dagger$} &0.21 &3.2 &2.5 &53 &$-0.5$ &$-1.0$ \\
J0024\textminus7204R &574.64 &4.0 &0.0057\tablenotemark{$\ddagger$} &0.29 &3.3 &2.5 &51 &$-0.3$ &$-0.8$ \\
J0024\textminus7204S &706.61 &4.0 &0.0045\tablenotemark{$\dagger$} &0.34 &2.6 &2.0 &77 &$-0.4$ &$-0.8$ \\
J0024\textminus7204T &263.56 &4.0 &0.011\tablenotemark{$\dagger$} &0.22 &12 &9.1 &19 &$-0.4$ &$-0.8$ \\
J0024\textminus7204U &460.53 &4.0 &0.0042\tablenotemark{$\dagger$} &0.37 &6.6 &5.1 &88 &$0.1$ &$-0.4$ \\
J0024\textminus7204W &850.22 &4.0 &0.0057\tablenotemark{$\ddagger$} &0.38 &2.0 &1.5 &66 &$-0.6$ &$-1.2$ \\
J0024\textminus7204X &419.15 &4.0 &0.0051\tablenotemark{$\dagger$} &0.47 &10 &7.8 &93 &$-0.2$ &$-0.5$ \\
J0024\textminus7204Y &910.47 &4.0 &0.0029\tablenotemark{$\dagger$} &0.42 &1.9 &1.5 &150 &$-0.4$ &$-0.7$ \\
J0024\textminus7204Z &439.13 &4.0 &0.0057\tablenotemark{$\ddagger$} &0.23 &4.5 &3.5 &40 &$-0.5$ &$-0.8$ \\
J0030+0451 &411.06 &0.3 &0.039 &0.27 &0.46 &0.36 &7.1 &$-0.6$ &$-0.8$ \\
J0034\textminus0534 &1065.43 &1.0 &0.013 &0.49 &0.40 &0.31 &36 &$-0.8$ &$-1.0$ \\
J0102+4839 &674.74 &4.0 &0.0039 &0.26 &2.2 &1.7 &67 &$-0.7$ &$-0.9$ \\
J0218+4232 &860.92 &3.1 &0.015 &0.42 &1.7 &1.3 &28 &$-0.5$ &$-0.8$ \\
J0340+4130 &606.18 &2.7 &0.0044 &0.31 &2.1 &1.7 &70 &$-0.5$ &$-0.9$ \\
J0348+0432 &51.12 &2.1 &0.0095 &0.48 &360 &280 &50 &$-0.8$ &$-0.9$ \\
J0407+1607 &77.82 &4.1 &0.0037 &0.22 &140 &110 &61 &$-0.6$ &$-1.0$ \\
J0437\textminus4715 &347.38 &0.2 &0.16 &0.22 &0.28 &0.21 &1.4 &$-1.0$ &$-1.1$ \\
J0453+1559 &43.69 &1.8 &0.0089 &0.66 &600 &460 &74 &$-0.8$ &$-1.0$ \\
J0605+37 &733.15 &1.2 &0.0091 &0.33 &0.67 &0.52 &36 &$-0.6$ &$-0.8$ \\
J0609+2130 &35.91 &1.8 &0.0091 &1.0 &1300 &1000 &110 &$-0.9$ &$-1.2$ \\
J0610\textminus2100 &517.96 &5.6 &0.0026 &0.29 &5.8 &4.5 &110 &$-0.7$ &$-0.9$ \\
J0613\textminus0200 &653.20 &1.1 &0.013 &0.37 &0.90 &0.70 &28 &$-0.3$ &$-0.7$ \\
J0614\textminus3329 &635.19 &1.0 &0.019 &0.46 &1.1 &0.85 &25 &$-0.2$ &$-0.7$ \\
J0621+1002 &69.31 &1.9 &0.0055 &0.30 &110 &85 &54 &$-0.8$ &$-1.0$ \\
J0636+5129 &697.12 &0.2 &0.044 &0.32 &0.13 &0.097 &7.3 &$-0.6$ &$-0.8$ \\
J0645+5158 &225.90 &0.8 &0.0078 &0.23 &3.3 &2.5 &30 &$-0.2$ &$-0.7$ \\
J0711\textminus6830 &364.23 &1.0 &0.013 &0.27 &2.0 &1.5 &21 &$-0.3$ &$-0.7$ \\
J0721\textminus2038 &128.68 &3.9 &0.0035 &0.40 &89 &68 &120 &$-0.2$ &$-0.7$ \\
J0737\textminus3039A &88.11 &1.1 &0.065 &0.26 &34 &27 &4.0 &$-0.6$ &$-0.9$ \\
J0742+66 &693.06 &0.9 &0.019 &0.40 &0.68 &0.53 &21 &$-0.6$ &$-0.8$ \\
J0751+1807 &574.92 &0.4 &0.030 &0.46 &0.53 &0.41 &15 &$0.3$ &$-0.6$ \\
J0900\textminus3144 &180.02 &0.8 &0.021 &0.89 &21 &17 &43 &$-0.6$ &$-1.0$ \\
J0908\textminus4913 &18.73 &1.0 &2.6 &14 &37000 &29000 &5.3 &$-0.7$ &$-0.9$ \\
J0931\textminus1902 &431.22 &3.6 &0.0020 &0.21 &4.0 &3.1 &110 &$-0.6$ &$-0.9$ \\
J0940\textminus5428 &22.84 &4.3 &1.2 &3.9 &30000 &23000 &3.4 &$-0.7$ &$-1.0$ \\
J1012+5307 &380.54 &0.7 &0.021 &0.36 &1.6 &1.3 &17 &$-0.0$ &$-0.8$ \\
J1016\textminus5819 &22.77 &4.6 &0.16 &3.2 &27000 &21000 &20 &$-0.7$ &$-1.2$ \\
J1016\textminus5857 &18.62 &9.3 &1.1 &14 &360000 &280000 &13 &$-0.9$ &$-1.1$ \\
J1017\textminus7156 &855.24 &0.7 &0.011 &0.36 &0.33 &0.25 &34 &$-0.7$ &$-0.9$ \\
J1022+1001 &121.56 &0.7 &0.017 &0.18 &8.5 &6.6 &11 &$-0.7$ &$-1.0$ \\
J1024\textminus0719 &387.43 &1.1 &0.014 &0.33 &2.3 &1.8 &24 &$-0.1$ &$-0.5$ \\
J1028\textminus5819 &21.88 &2.8 &1.2 &9.7 &53000 &41000 &7.9 &$0.6$ &$-0.6$ \\
J1038+0032 &69.32 &2.4 &0.0052 &0.29 &130 &100 &56 &$-0.9$ &$-1.0$ \\
J1045\textminus4509 &267.59 &0.3 &0.036 &0.19 &0.87 &0.67 &5.3 &$-0.4$ &$-0.8$ \\
J1055\textminus6028 &20.07 &30.0 &0.16 &11 &800000 &620000 &72 &$-1.1$ &$-1.3$ \\
J1105\textminus6107 &31.65 &7.1 &0.60 &1.8 &12000 &9200 &3.0 &$-0.8$ &$-1.0$ \\
J1112\textminus6103 &30.78 &30.0 &0.19 &1.5 &44000 &34000 &7.8 &$-1.7$ &$-1.7$ \\
J1122+78 &476.01 &0.6 &0.016 &0.23 &0.61 &0.47 &14 &$-0.6$ &$-0.9$ \\
J1125\textminus6014 &760.35 &1.9 &0.0045 &0.53 &1.7 &1.3 &120 &$-0.2$ &$-0.6$ \\
J1142+0119 &394.07 &2.0 &0.0068 &0.29 &3.7 &2.8 &43 &$-0.3$ &$-0.6$ \\
J1231\textminus1411 &542.91 &0.5 &0.044 &0.30 &0.43 &0.33 &6.7 &$-0.6$ &$-0.8$ \\
J1300+1240 &321.62 &0.6 &0.058 &0.22 &1.2 &0.94 &3.9 &$-0.6$ &$-0.9$ \\
J1302\textminus3258 &530.38 &1.9 &0.0057 &0.22 &1.4 &1.0 &38 &$-0.7$ &$-1.0$ \\
J1312+0051 &473.03 &1.1 &0.014 &0.26 &1.3 &0.99 &18 &$-0.5$ &$-1.1$ \\
J1327\textminus0755 &746.85 &2.2 &0.0031 &0.38 &1.4 &1.1 &120 &$-0.5$ &$-0.8$ \\
J1410\textminus6132 &39.96 &30.0 &0.21 &0.77 &14000 &11000 &3.6 &$-1.0$ &$-1.1$ \\
J1418\textminus6058 &18.08 &1.6 &6.2 &14 &63000 &49000 &2.2 &$0.2$ &$-0.5$ \\
J1446\textminus4701 &911.29 &2.0 &0.0082 &0.45 &1.0 &0.80 &54 &$-0.4$ &$-1.1$ \\
J1453+1902 &345.29 &0.9 &0.012 &0.32 &2.4 &1.8 &26 &$-0.7$ &$-0.8$ \\
J1455\textminus3330 &250.40 &0.7 &0.019 &0.21 &2.3 &1.8 &11 &$-0.5$ &$-0.8$ \\
J1509\textminus5850 &22.49 &3.9 &0.67 &2.5 &18000 &14000 &3.7 &$-1.0$ &$-1.1$ \\
J1518+4904 &48.86 &0.7 &0.0094 &0.37 &100 &79 &39 &$-1.1$ &$-1.2$ \\
J1524\textminus5625 &25.57 &3.8 &1.5 &4.0 &22000 &17000 &2.7 &$-0.5$ &$-0.8$ \\
J1531\textminus5610 &23.75 &3.1 &1.1 &3.3 &17000 &13000 &3.1 &$-0.6$ &$-1.1$ \\
J1537+1155 &52.76 &1.1 &0.061 &0.46 &160 &130 &7.4 &$-0.3$ &$-0.7$ \\
J1545\textminus4550 &559.40 &2.0 &0.015 &0.27 &1.6 &1.3 &18 &$-0.6$ &$-0.9$ \\
J1551\textminus0658 &281.94 &1.5 &0.0094 &0.27 &4.7 &3.6 &28 &$-0.3$ &$-1.0$ \\
J1600\textminus3053 &555.88 &1.8 &0.0073 &0.27 &1.5 &1.2 &38 &$-0.6$ &$-1.0$ \\
J1603\textminus7202 &134.75 &0.5 &0.016 &0.18 &5.0 &3.8 &11 &$-0.3$ &$-0.9$ \\
J1614\textminus2230 &634.76 &0.7 &0.020 &0.51 &0.84 &0.65 &25 &$0.4$ &$-0.1$ \\
J1618\textminus3921 &166.84 &4.8 &0.0036 &0.19 &30 &23 &52 &$-0.4$ &$-0.8$ \\
J1623\textminus2631 &180.57 &1.8 &0.013\tablenotemark{$\ddagger$} &0.27 &14 &11 &21 &$-0.3$ &$-0.7$ \\
J1630+37 &602.75 &0.8 &0.017 &0.31 &0.68 &0.53 &18 &$-0.5$ &$-0.8$ \\
J1640+2224 &632.25 &1.4 &0.0053 &0.54 &1.9 &1.4 &100 &$0.3$ &$-0.3$ \\
J1643\textminus1224 &432.75 &0.7 &0.022 &0.27 &1.0 &0.77 &12 &$-0.3$ &$-0.7$ \\
J1653\textminus2054 &484.36 &2.6 &0.0050 &0.26 &2.7 &2.1 &51 &$-0.5$ &$-0.9$ \\
J1708\textminus3506 &443.94 &3.5 &0.0037 &0.25 &4.2 &3.2 &68 &$-0.5$ &$-0.9$ \\
J1709+2313 &431.85 &1.8 &0.0039 &0.30 &2.8 &2.1 &76 &$-0.6$ &$-0.9$ \\
J1709\textminus4429 &19.51 &2.6 &3.0 &6.1 &40000 &31000 &2.1 &$-0.7$ &$-1.0$ \\
J1709\textminus4429\tablenotemark{*} &19.51 &2.6 &3.0 &4.6 &30000 &23000 &1.5 &$-0.8$ &$-1.0$ \\
J1710+49 &621.07 &0.4 &0.049 &0.28 &0.27 &0.21 &5.8 &$-0.6$ &$-1.0$ \\
J1713+0747 &437.62 &1.2 &0.0093 &0.36 &2.1 &1.6 &38 &$0.1$ &$-0.4$ \\
J1718\textminus3825 &26.78 &4.2 &0.80 &1.7 &9800 &7500 &2.2 &$-0.8$ &$-1.1$ \\
J1719\textminus1438 &345.41 &1.6 &0.0058 &0.24 &3.1 &2.4 &42 &$-1.0$ &$-1.1$ \\
J1721\textminus2457 &571.98 &1.6 &0.0065 &0.32 &1.4 &1.1 &49 &$-0.5$ &$-0.8$ \\
J1727\textminus2946 &73.85 &1.6 &0.015 &0.29 &82 &63 &19 &$-0.8$ &$-1.1$ \\
J1729\textminus2117 &30.17 &1.4 &0.0093 &1.0 &1500 &1100 &110 &$-1.0$ &$-1.1$ \\
J1730\textminus2304 &246.22 &0.6 &0.020 &0.17 &1.6 &1.3 &8.2 &$-0.7$ &$-1.1$ \\
J1731\textminus1847 &853.04 &4.0 &0.0066 &0.56 &2.9 &2.3 &84 &$-0.5$ &$-0.7$ \\
J1732\textminus5049 &376.47 &1.8 &0.0073 &0.18 &2.1 &1.6 &24 &$-0.7$ &$-0.9$ \\
J1738+0333 &341.87 &1.5 &0.011 &0.23 &2.8 &2.1 &21 &$-0.9$ &$-1.1$ \\
J1741+1351 &533.74 &1.1 &0.021 &0.50 &1.8 &1.4 &24 &$0.5$ &$-0.2$ \\
J1744\textminus1134 &490.85 &0.4 &0.030 &0.40 &0.63 &0.49 &13 &$-0.0$ &$-0.5$ \\
J1745+1017 &754.11 &1.4 &0.0063 &0.45 &1.0 &0.79 &71 &$0.1$ &$-0.4$ \\
J1745\textminus0952 &103.22 &2.4 &0.0074 &0.22 &46 &35 &29 &$-0.7$ &$-0.9$ \\
J1748\textminus2446A &172.96 &5.5 &0.0041\tablenotemark{$\ddagger$} &0.21 &36 &28 &50 &$-0.4$ &$-0.9$ \\
J1748\textminus3009 &206.53 &6.0 &0.0026 &0.21 &27 &21 &79 &$-0.4$ &$-0.8$ \\
J1750\textminus2536 &57.55 &3.5 &0.0035 &0.41 &410 &320 &120 &$-0.8$ &$-1.0$ \\
J1751\textminus2857 &510.87 &1.4 &0.0095 &0.52 &2.7 &2.1 &55 &$-0.1$ &$-0.6$ \\
J1753\textminus1914 &31.77 &2.8 &0.016 &1.3 &3400 &2600 &80 &$-0.5$ &$-0.7$ \\
J1753\textminus2240 &21.02 &3.5 &0.0071 &5.5 &40000 &31000 &770 &$-0.8$ &$-1.0$ \\
J1756\textminus2251 &70.27 &0.7 &0.066 &0.36 &50 &39 &5.4 &$-0.5$ &$-1.0$ \\
J1757\textminus27 &113.08 &5.4 &0.0012 &0.34 &140 &100 &290 &$-0.2$ &$-0.5$ \\
J1801\textminus1417 &551.71 &1.8 &0.0054 &0.38 &2.1 &1.6 &69 &$-0.4$ &$-0.7$ \\
J1801\textminus3210 &268.33 &5.1 &0.000046 &0.25 &17 &13 &5500 &$-0.3$ &$-0.8$ \\
J1802\textminus2124 &158.13 &3.3 &0.0058 &0.17 &21 &16 &29 &$-0.7$ &$-0.9$ \\
J1804\textminus0735 &86.58 &7.8 &0.0029\tablenotemark{$\ddagger$} &0.28 &280 &210 &97 &$-0.7$ &$-0.9$ \\
J1804\textminus2717 &214.06 &1.2 &0.014 &0.23 &5.5 &4.3 &16 &$0.2$ &$-0.5$ \\
J1810+1744 &1202.82 &2.5 &0.0053 &0.49 &0.80 &0.61 &92 &$-0.6$ &$-0.9$ \\
J1811\textminus2405 &751.71 &1.7 &0.011 &0.30 &0.86 &0.66 &28 &$-0.7$ &$-0.9$ \\
J1813\textminus2621 &451.47 &3.4 &0.0040 &0.24 &3.7 &2.9 &60 &$-0.5$ &$-0.8$ \\
J1823\textminus3021A &367.65 &8.6 &0.023 &0.27 &23 &17 &11 &$-0.3$ &$-0.9$ \\
J1824\textminus2452A &654.81 &5.1 &0.036 &0.45 &5.5 &4.2 &12 &$-0.0$ &$-0.5$ \\
J1825\textminus0319 &439.22 &3.3 &0.0031 &0.26 &4.1 &3.2 &83 &$-0.5$ &$-0.8$ \\
J1826\textminus1334 &19.71 &4.1 &1.7 &12 &120000 &93000 &7.2 &$-0.3$ &$-0.7$ \\
J1828\textminus1101 &27.76 &7.3 &0.50 &3.3 &30000 &23000 &6.7 &$-0.6$ &$-0.9$ \\
J1832\textminus0836 &735.53 &1.4 &0.010 &0.34 &0.83 &0.64 &34 &$-0.6$ &$-0.9$ \\
J1833\textminus0827 &23.45 &4.5 &0.62 &17 &130000 &100000 &28 &$12.4$ &$-5.5$ \\
J1837\textminus0604 &20.77 &6.2 &0.89 &7.9 &110000 &83000 &8.8 &$-0.4$ &$-0.7$ \\
J1840\textminus0643 &56.21 &6.7 &0.0020 &0.33 &660 &510 &160 &$-1.0$ &$-1.1$ \\
J1843\textminus1113 &1083.62 &2.0 &0.0094 &0.46 &0.73 &0.56 &49 &$-0.7$ &$-1.0$ \\
J1845\textminus0743 &19.10 &5.8 &0.082 &6.5 &98000 &76000 &79 &$-0.8$ &$-1.0$ \\
J1853+1303 &488.78 &1.6 &0.0074 &0.32 &2.0 &1.6 &43 &$-0.6$ &$-0.9$ \\
J1853\textminus0004 &19.72 &6.6 &0.29 &18 &280000 &220000 &62 &$-0.2$ &$-0.6$ \\
J1856+0245 &24.72 &10.3 &0.69 &2.6 &41000 &32000 &3.7 &$-0.8$ &$-1.0$ \\
J1857+0943 &372.99 &0.7 &0.021 &0.51 &2.4 &1.9 &24 &$0.3$ &$-0.6$ \\
J1903+0327 &930.27 &6.5 &0.0037 &0.41 &2.9 &2.2 &110 &$-0.8$ &$-1.0$ \\
J1903\textminus7051 &555.88 &1.1 &0.029 &0.42 &1.4 &1.1 &14 &$-0.3$ &$-0.8$ \\
J1909\textminus3744 &678.63 &1.1 &0.015 &0.28 &0.65 &0.50 &18 &$-0.6$ &$-0.9$ \\
J1910+1256 &401.32 &1.9 &0.0058 &0.30 &3.4 &2.7 &52 &$-0.6$ &$-0.9$ \\
J1910\textminus5959A &612.33 &4.5 &0.0051\tablenotemark{$\ddagger$} &0.26 &2.9 &2.2 &50 &$0.0$ &$-0.5$ \\
J1910\textminus5959C &378.98 &4.5 &0.0051\tablenotemark{$\ddagger$} &0.21 &6.3 &4.9 &42 &$-0.4$ &$-1.2$ \\
J1910\textminus5959D &221.35 &4.5 &0.0051\tablenotemark{$\ddagger$} &0.16 &14 &11 &32 &$-0.3$ &$-0.8$ \\
J1911+1347 &432.34 &1.6 &0.0096 &0.29 &2.4 &1.8 &30 &$-0.4$ &$-0.7$ \\
J1911\textminus1114 &551.61 &1.6 &0.010 &0.27 &1.3 &1.0 &27 &$-0.6$ &$-1.2$ \\
J1915+1606 &33.88 &7.1 &0.014 &0.86 &5000 &3900 &63 &$-0.8$ &$-1.0$ \\
J1918\textminus0642 &261.58 &0.9 &0.016 &0.16 &2.1 &1.6 &10 &$-0.7$ &$-0.9$ \\
J1923+2515 &527.96 &1.0 &0.013 &0.25 &0.83 &0.64 &19 &$-0.7$ &$-1.1$ \\
J1925+1721 &26.43 &9.6 &0.31 &3.0 &38000 &30000 &9.5 &$-0.6$ &$-0.7$ \\
J1932+17 &47.81 &2.7 &0.023 &1.3 &1400 &1100 &54 &$1.6$ &$0.9$ \\
J1935+2025 &24.96 &8.6 &0.81 &3.3 &44000 &34000 &4.1 &$-0.6$ &$-1.0$ \\
J1939+2134 &1283.86 &1.5 &0.044 &0.48 &0.42 &0.32 &11 &$-0.7$ &$-1.0$ \\
J1943+2210 &393.38 &8.3 &0.0013 &0.26 &13 &10 &200 &$-0.4$ &$-0.8$ \\
J1944+0907 &385.71 &1.3 &0.012 &0.30 &2.4 &1.9 &26 &$-0.4$ &$-0.7$ \\
J1946+3417 &630.89 &6.4 &0.0013 &0.32 &4.9 &3.8 &260 &$-0.5$ &$-0.8$ \\
J1949+3106 &152.23 &7.8 &0.0028 &0.20 &64 &49 &72 &$-0.5$ &$-0.8$ \\
J1950+2414 &464.60 &7.3 &0.0023 &0.36 &11 &8.9 &150 &$-0.2$ &$-0.7$ \\
J1955+2527 &410.44 &9.1 &0.0012 &0.35 &18 &14 &280 &$-0.0$ &$-0.6$ \\
J1955+2908 &326.10 &5.4 &0.0033 &0.19 &9.2 &7.1 &58 &$-0.7$ &$-0.9$ \\
J1959+2048 &1244.24 &1.5 &0.017 &0.74 &0.69 &0.54 &44 &$-0.4$ &$-0.8$ \\
J2007+2722 &81.64 &6.8 &0.0074 &0.39 &380 &290 &53 &$-0.4$ &$-0.9$ \\
J2010\textminus1323 &382.90 &1.3 &0.0060 &0.28 &2.3 &1.8 &46 &$-0.5$ &$-0.8$ \\
J2017+0603 &690.56 &1.3 &0.010 &0.46 &1.2 &0.92 &45 &$-0.4$ &$-1.1$ \\
J2019+2425 &508.32 &0.9 &0.012 &0.55 &1.8 &1.4 &46 &$-0.1$ &$-0.7$ \\
J2033+1734 &336.19 &1.4 &0.0081 &0.27 &3.1 &2.4 &34 &$-0.6$ &$-1.0$ \\
J2043+1711 &840.38 &1.2 &0.0096 &0.32 &0.53 &0.41 &33 &$-0.6$ &$-1.0$ \\
J2047+1053 &466.64 &2.2 &0.0080 &0.21 &2.0 &1.6 &26 &$-0.7$ &$-0.9$ \\
J2051\textminus0827 &443.59 &1.3 &0.011 &0.21 &1.3 &1.0 &20 &$-0.7$ &$-0.9$ \\
J2124\textminus3358 &405.59 &0.4 &0.041 &0.29 &0.68 &0.52 &7.1 &$-0.3$ &$-1.0$ \\
J2129+1210A &18.07 &12.9 &0.0018\tablenotemark{$\ddagger$} &14 &530000 &410000 &8000 &$-0.6$ &$-0.8$ \\
J2129+1210B &35.63 &12.9 &0.0018\tablenotemark{$\ddagger$} &0.90 &8600 &6700 &510 &$-1.1$ &$-1.2$ \\
J2129+1210C &65.51 &10.0 &0.010\tablenotemark{$\dagger$} &0.34 &970 &750 &33 &$-0.8$ &$-1.0$ \\
J2129+1210D &416.42 &12.9 &0.0018\tablenotemark{$\ddagger$} &0.23 &16 &12 &130 &$-0.6$ &$-0.9$ \\
J2129+1210E &429.97 &12.9 &0.0018\tablenotemark{$\ddagger$} &0.34 &22 &17 &190 &$-0.1$ &$-0.5$ \\
J2129\textminus5721 &536.72 &3.2 &0.0060 &0.28 &2.9 &2.3 &47 &$0.1$ &$-0.4$ \\
J2145\textminus0750 &124.59 &0.5 &0.021 &0.19 &6.2 &4.8 &9.2 &$-0.4$ &$-0.8$ \\
J2214+3000 &641.18 &1.0 &0.018 &0.56 &1.3 &1.0 &32 &$-0.1$ &$-0.6$ \\
J2222\textminus0137 &60.94 &0.3 &0.040 &0.30 &21 &16 &7.5 &$-1.1$ &$-1.2$ \\
J2229+2643 &671.63 &1.4 &0.0040 &0.39 &1.2 &0.90 &96 &$-0.4$ &$-0.9$ \\
J2234+06 &559.19 &1.1 &0.013 &0.32 &1.1 &0.85 &25 &$-0.5$ &$-0.8$ \\
J2235+1506 &33.46 &1.1 &0.011 &0.80 &780 &600 &71 &$-1.0$ &$-1.1$ \\
J2241\textminus5236 &914.62 &0.7 &0.021 &0.45 &0.35 &0.27 &21 &$-0.5$ &$-0.9$ \\
J2302+4442 &385.18 &0.8 &0.018 &0.19 &0.89 &0.69 &11 &$-0.1$ &$-0.4$ \\
J2317+1439 &580.51 &1.9 &0.0036 &0.40 &2.1 &1.6 &110 &$-0.2$ &$-0.6$ \\
J2322+2057 &415.94 &0.8 &0.015 &0.25 &1.1 &0.84 &17 &$-0.6$ &$-0.8$
\enddata
\tablenotetext{$\dagger$}{The pulsar's spin-down is corrected for proper motion effects.}\\
\tablenotetext{$\ddagger$}{The pulsar's spin-down is calculated using a characteristic spin-down age of 
$10^9$ years and a braking index, $n$, of 5 (i.e.\ braking due to gravitational radiation).}
\tablenotetext{*}{Uses a restricted prior on orientation parameters (see Appendix~\ref{app:restricted}).}

\tablecomments{This does not include the high-value targets already listed in 
Table~\ref{tab:highvalueres}. For PSR\,J0023+0923 and PSR\,J0340+4130, instrinsic period derivatives are available in the 
ATNF pulsar catalog (v.~1.54) \citep{Manchester:2005}; however, they are incorrect and therefore the 
spin-down limits have been calculated using the observed spin-down. For eight pulsars in the globular 
cluster 47 Tuc (PSRs~J0024\tmin7204E, H, Q, S, T, U, X, and Y) we have obtained (P.~C.~C.~Freire 2016, private communication) intrinsic period 
derivatives to calculate the spin-down limits, with that for X being the $3\sigma$ upper limit from 
\citet{2016MNRAS.462.2918R} given that it gives a characteristic age older than $10^9$ years. For 
PSR\,J1823\textminus3021A (in globular cluster NGC~6624) and PSR\,J1824\textminus2452A (in globular cluster 
M28), we follow \citet{2011Sci...334.1107F} and \citet{2013ApJ...778..106J} and calculate the spin-down 
limit assuming that the contributions to the observed $\dot{f}_{\rm rot}$ are negligably affected by cluster 
accelerations. The intrinsic spin-down for PSR\,J2129+1210C (in globular cluster M15) is taken from 
\citet{2004ApJ...602..264M}, which shows that the observed spin-down is negligably affected by 
accelerations \citep[it is in the outskirts of the cluster as is shown in][]{Anderson:1993}. The following 
pulsars use distance estimates that are not taken from the values given in the ATNF pulsar catalog: 
PSR\,J1017\textminus7156 (updated parallax distance provided by R.~M.~Shannon 2016, private communication), PSR\,J1418\textminus6058 
\citep[distance to more distant association in][]{1997ApJ...476..347Y}, PSR\,J1813\textminus1246 (lower 
limit on distance from \citet{2014ApJ...795..168M}), PSR\,J1823\textminus3021A \citep[distance for NGC~6624 
in][]{2007AJ....133.1287V}, PSR\,J1824\textminus2452A \citep[distance for M28 in][]{1991AJ....102..152R}, 
PSR\,J1826\textminus1256 \citep[lower distance range from][]{Wang:2011, 2016MNRAS.458.2813V}, 
PSRs\,J1910\tmin5959A, C, and D \citep[distances of 4.45\,kpc calculated from the distance modulus to NGC~6752 
in Table~4 of][]{2003A&A...408..529G}, PSR\,J2129+1210C \citep{2004ApJ...602..264M}, and PSR\,J2234+06 
(P.~C.~C.~Freire 2016, private communication).}

\end{deluxetable*}

\bibliography{O1KnownPulsarPaper}

\end{document}